\documentclass[twocolumn,pra,aps,superscriptaddress]{revtex4-2}
\usepackage{graphicx}
\usepackage{braket}
\usepackage{amsmath,amsfonts,amssymb,amstext,amscd,amsthm}
\usepackage{comment}
\usepackage[font=small,format=plain,labelsep=period,labelfont=normal,textfont=normal,singlelinecheck=false,justification=raggedright]{caption}
\usepackage[colorlinks, citecolor=blue, urlcolor=red, linkcolor=blue]{hyperref}
\usepackage{dsfont}
\usepackage{xcolor}

\definecolor{mango}{rgb}{1.0, 0.75, 0.04}
\definecolor{fire}{rgb}{0.98, 0.34,  0.03}
\definecolor{hotpink}{rgb}{1.0, 0.0, 0.43}
\definecolor{violetflower}{rgb}{0.51, 0.22, 0.93}
\definecolor{skyblue}{rgb}{0.23, 0.53, 1.0}

\colorlet{MANGO}{mango}
\colorlet{FIRE}{fire}
\colorlet{HOTPINK}{hotpink}
\colorlet{VIOLETFLOWER}{violetflower}
\colorlet{SKYBLUE}{skyblue}

\usepackage[normalem]{ulem}

\begin{document}
\title{Multi-qubit quantum computing using discrete-time quantum walks on closed graphs}
\author{Prateek Chawla}
\affiliation{The Institute of Mathematical Sciences, C. I. T. Campus, Taramani, Chennai 600113, India}
\affiliation{Homi Bhabha National Institute, Training School Complex, Anushakti Nagar, Mumbai 400094, India}
\email{prateekc@imsc.res.in}
\author{Shivani Singh}
\affiliation{The Institute of Mathematical Sciences, C. I. T. Campus, Taramani, Chennai 600113, India}
\affiliation{Homi Bhabha National Institute, Training School Complex, Anushakti Nagar, Mumbai 400094, India}
\author{Aman Agarwal}
\affiliation{BITS-Pilani, K. K. BIrla Goa Campus, NH17B, Bypass Road, Zuarinagar, Goa 403726, India}
\author{Sarvesh Srinivasan}
\affiliation{Birla Institute of Technology and Science, Pilani, Pilani Campus, Pilani 333031, India}
\author{C. M. Chandrashekar}
\email{chandru@imsc.res.in}
\affiliation{The Institute of Mathematical Sciences, C. I. T. Campus, Taramani, Chennai 600113, India}
\affiliation{Homi Bhabha National Institute, Training School Complex, Anushakti Nagar, Mumbai 400094, India}

\begin{abstract}

Universal quantum computation can be realised using both continuous-time and discrete-time quantum walks.  We present a version based on single particle discrete-time quantum walk to realize multi-qubit computation tasks. The scalability of the scheme is demonstrated by using a set of walk operations on a closed lattice form to implement the universal set of quantum gates on multi-qubit system.  We also present a set of experimentally realizable walk operations that can implement Grover's algorithm, quantum Fourier transformation and quantum phase estimation algorithms. An elementary implementation of error detection and correction is also presented. Analysis of space and time complexity  of the scheme highlights the advantages of quantum walk based model for quantum computation on systems where implementation of quantum walk evolution operations is an inherent feature of the system. 

\end{abstract}

\maketitle
\section{\label{sec:intro}Introduction}

\noindent
Quantum computing is poised to provide supremacy over classical computing using quantum mechanical phenomena such as superposition, interference and entanglement. Physical systems like, superconducting circuits\,\cite{JF08, WST04}, nuclear magnetic resonance (NMR) systems\,\cite{CMR09, J00, LR13, SA08}, ion traps \cite{P90,BDCD04}, ultra-cold atoms in optical lattice\,\cite{J04, GB17} and photonics\,\cite{ERG01,SS99,PWK07}  have been successfully engineered to demonstrate small scale quantum processors and implement quantum simulations and computational tasks.  The noisy-intermediate scale quantum processors we have today are still far from the one that can be used for performing useful tasks that are inaccessible by the existing powerful classical computers. Different models for quantum computation and the engineering of different physical systems and architecture to build scalable processors has been explored for a long time now. For example, measurement based quantum computing model\,\cite{RDH03,PKT05,TPR14}, adiabatic quantum computing model\,\cite{EJS00,GE06,JP08}, and KLM-linear optical quantum computing\,\cite{KLM01} are some of the examples in addition to standard circuit based quantum computation model. The use of quantum walks\,\cite{meyer1996quantum, kempe2003quantum, CSL08, venegas2012quantum}, which are part of several quantum algorithms\,\cite{Amb03} developed to outperform classical algorithms at computational tasks has also been proposed to develop a scheme for universal quantum computation model.  

Quantum walk based quantum computing model was first introduced on unweighted graph using the continuous-time quantum walk\,\cite{AMC09} and a corresponding scheme using discrete-time quantum walk was later proposed\,\cite{LCE10}. Recently, we proposed a new scheme using a single qubit discrete-time quantum walk on a closed lattice setting\,\cite{SPAC21}. Compared to the earlier discrete-time quantum walk scheme which requires large number of real qubits and higher dimensional coin operation, our scheme defines computation purely as a sequence of position dependent coin and shift operations on a system with single real qubit and position space as a additional computational basis. Therefore, our scheme is less resource-intensive and can be physically realizable on any lattice based systems. With the advent of photonics-based quantum computing systems \cite{ZWDC20} and the efficient realization of quantum walks \cite{ZLLL10}, the potential realizability of our proposed scheme gains more relevance.

Here we present a detailed extension of the simple, implementable quantum computing scheme using a single particle discrete-time quantum walk which can be scaled to higher dimensions\,\cite{SPAC21}. Along with the position Hilbert space on which the quantum walks are defined, the discrete-time quantum walk provides additional degree of freedom in the form of coin Hilbert space that can be exploited to achieve control over the states to perform computing operations. This model can be implemented on a photonic or lattice based quantum systems where one photon or free particle can act as coin that can be used to perform computation when entangled with the position space. We propose the use of multiple sets of closed graph with four sites and four edges to act as a system with {$2^{(N-1)}$-dimensional position space}. Each graph is equivalent to two-qubit state and $n$-sets of closed graph provides 2n-qubit equivalent states. With the help of the coin and shift operations, the particle (coin) and the position state can be evolved into the desired output state\,\cite{GPE19}. We  also demonstrate the effectiveness of our scheme by presenting  a combination of quantum walk operations to implement the quantum algorithms like Grover's search algorithm, quantum Fourier transformation and phase estimation algorithms. Further, an elementary implementation of single qubit error detection (3-qubit code) for both bit-flip and phase-flip errors, and error correction using a 5-qubit code is presented. We also discuss the space and time complexity of the scheme in a generic sense to highlight the possible advantages of the quantum walk based scheme.  

In section \ref{sec:dtqw} we present a brief discussion on the discrete-time quantum walk and show the scalability of the single qubit quantum computational scheme to N-qubit equivalent system by expanding the position space. Section \ref{sec:gates} shows the implementation of universal gates on this N-qubit equivalent system, and in sections \ref{sec:grover}, \ref{sec:qft} and \ref{sec:qpe} we present schemes for realization of Grover's search algorithm, quantum Fourier transformation and phase estimation algorithm on the DTQW-based system, respectively. We discuss the space and time complexity of quantum walk based scheme in section \ref{sec:complexity}, and explore a basic implementation of quantum error detection and correction codes in section \ref{sec:error}. We present our conclusions and future outlook for this work in section \ref{sec:conc}.

\section{\label{sec:dtqw} {Quantum walk on higher-qubit equivalent systems}}
\noindent
The dynamics of the discrete-time quantum walk on a closed graph is defined on a Hilbert space $\mathcal{H} = \mathcal{H}_c \otimes \mathcal{H}_p$ where, $\mathcal{H}_c$ is the coin Hilbert space with internal degrees of freedom and $\mathcal{H}_p$ is the position Hilbert space defined by closed set of points in the position space\,\cite{BSC08}. For the computation model proposed in this work, we choose the position Hilbert space to be defined by 
the multiple sets of closed graphs of 4-states spanned by $ \ket{x} = \{ \ket{0}, \ket{1}, \ket{2}, \ket{3} \}$. The evolution operation on this setup of discrete-time quantum walk is described by the action of the unitary quantum coin operation $\hat{C}$ on the coin state followed by the conditional position shift operation on the desired set of closed graph of the position space. 

The general form of position shift operator for  discrete-time quantum walk on a closed graph, that translates to the left or right conditioned on the coin states with {$\mu$ internal degrees of freedom} is given as,
\begin{equation}
\label{eq:eq1}
\hat{S}_{\pm}^{\alpha}  =  \sum_{l\in\mathbb{Z}}  \Big [\ket{\alpha}\bra{\alpha}
\otimes   \ket{l\pm1 \mod 4}\bra{l}+ \sum_{\beta \neq \alpha}^{\mu}\Big(\ket{\beta}\bra{\beta} \otimes
\ket{l}\bra{l} \Big) \Big ].
\end{equation}
Here, $\{\ket{\alpha},\ket{\beta}\} \in \mathcal{H}_c$ are the basis states of coin Hilbert space $\mathcal{H}_c$ and $\ket{l}$ are the basis states of position Hilbert space $\mathcal{H}_p$.
The general form of the quantum coin operator with two internal degree of freedom $\mathcal{H}_c  = span \{\ket{0},  \ket{1}\}$ is given by SU(2) operator of the form, 
\begin{equation}
\label{eq:eq2}
\hat{C}(\xi,\zeta,\theta) =
  \begin{bmatrix}
   ~~e^{i \xi}\cos(\theta) & ~~~~ e^{i \zeta}\sin(\theta) \\
   e^{-i \zeta}   \sin(\theta) & - e^{-i \xi} \cos(\theta) 
  \end{bmatrix}.
\end{equation}
This set of operators along with the identity operator $\mathbb{I}$ can be considered a generic set of operators that describes the scalable quantum computation scheme using discrete-time quantum walk, hereafter called the quantum walk in this text. 
\vskip 0.1in
\noindent
\subsection*{{Quantum computation using quantum walk }}
The scheme presented for universal quantum computation on quantum walk for three qubit equivalent system \cite{SPAC21} can be scaled to a larger qubit system by using the same coin in conjunction with different sets of closed graph of the position space. This method will expand the shift operator with the increase of the number of closed graphs of four-sites, but can be {scaled}  as far as the scheme goes. 

\begin{figure}
	\centering
	\includegraphics[width=0.75\linewidth]{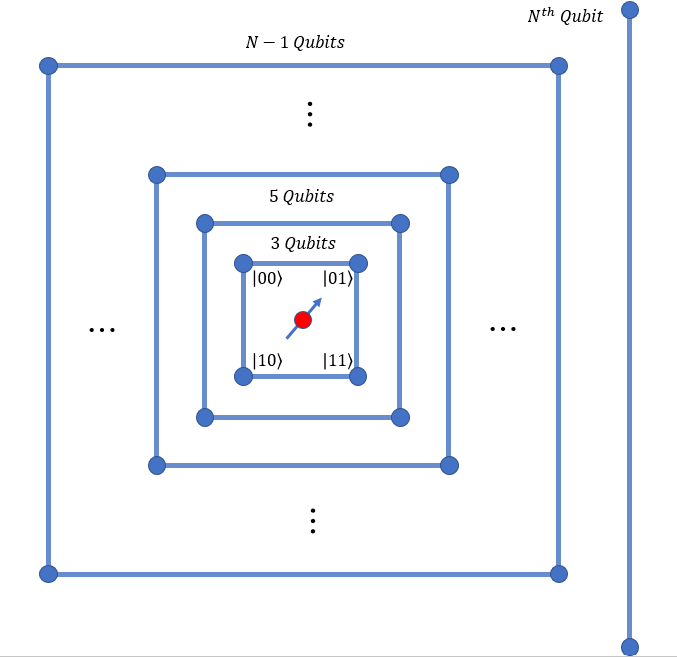}
	\caption{Scaling of the Quantum walk scheme to $N-$qubit system when $N$ is even to implement universal gates. It consist of $(N/2 -1)$ number of quantum walk system with four position states and one quantum walk with two position states.}
	\label{fig:Scaling_even}
\end{figure}

\begin{figure}
	\centering
	\includegraphics[width=0.75\linewidth]{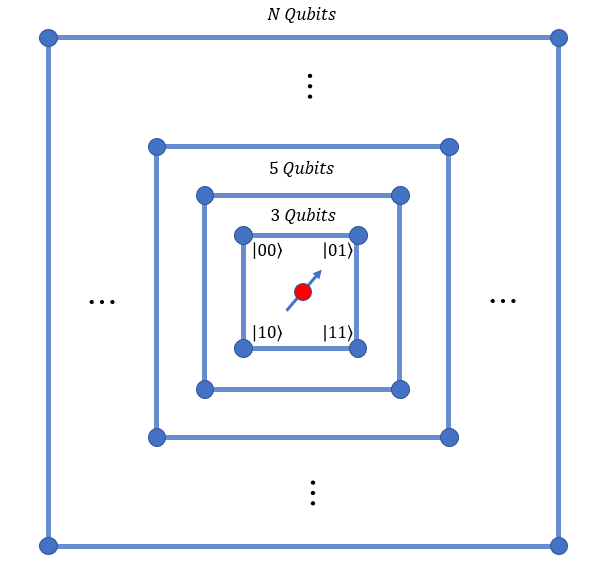}
	\caption{Scaling of the Quantum walk scheme to $N-$qubit system when $N$ is odd to implement universal gates. It consist of $(N-1)/2$ number of quantum walk system with four position states.}
	\label{fig:Scaling_odd}
\end{figure}

The form of shift operators which is used throughout for scaling of the universal computation model for input state $\ket{k}\bigotimes_{i=1}^{n} \ket{m_i}$ will be given as,
\begin{equation}
\label{eq:eq3}
\begin{split}
	S^k_{j,\pm} = \sum_l &\Big [ \ket{k}\bra{k}\otimes \mathbb{I}^{\otimes j-1} \otimes \ket{l\pm 1 \mod 4}\bra{l} \otimes \mathbb{I}^{\otimes n-j} \\
	+  &\ket{p\neq k}\bra{p}\otimes \mathbb{I}^{\otimes n} \Big ], 
\end{split}
\end{equation}
where, $n$ is the total number of closed graphs and $j$ indicates the closed graph on which the shift operation is performed. $\{\ket{k}, \ket{p}\} \in \mathcal{H}_c$ are states in the coin Hilbert space with two internal degree of freedom and $\ket{l}$ represents the {four states on the four-site closed graph,} and number of closed graph is $n$. The number of states for this case will be equivalent to the number of states in the combined state of the Hilbert-space $\mathcal{H}_c \otimes \mathcal{H}_{p}$, where $\mathcal{H}_p$ has dimension $2^{(N-1)}$, and $N$ is the total number of qubits in the system. The evolution operation on this system can be interpreted as the shift operation on the  $j^\text{th}$ closed graph representing the 'selected` position space and identity operation on the rest of the closed sets of the position space, as shown in Figs. \ref{fig:Scaling_even} and \ref{fig:Scaling_odd}.

This can be then used to derive the $\hat{W}$ operator,  in order to implement the Hadamard gate on  $N$-qubit system and a specific case of this $\hat{W}$-operator is used in Ref. \cite{SPAC21}. {An $N$-qubit system will require $n= \big[ \frac{(N-2)}{2} \big]$ sets of four-site closed graph and one set of two-site graph with one edge if $N$ is even, and $n= \big[ \frac{(N-1)}{2} \big]$ sets of four-site closed graphs if $N$ is odd.}  In order to simplify notation, we choose $\ket{m_j}$ to represent the position state of the $j^\text{th}$ set of closed graphs. The complete state of the position space is given by $\ket{m}$, which is defined as,
\begin{equation}
\label{eq:eq4}
\ket{m} \equiv \bigotimes_{i=1}^{n} \ket{m_i}.
\end{equation}
Then, the $\hat{W}$ operators on state $\ket{m_j}$ with $1<j<n$ is defined as,
\begin{align} \label{eq:eq5}
\hat{W}_{j,\pm}^0 \ket{k}\ket{m} &= \left. \Big[ ( \sigma_x \otimes \ket{m}\bra{m} + \mathbb{I}\otimes \sum_{l \neq m} \ket{l}\bra{l}) \right. \nonumber \\ 
&\left. S^0_{j,\pm} \left(\sigma_x \otimes \mathbb{I}^{\otimes n} \right) \right. \Big] \ket{k}\ket{m}, 
\end{align}
\begin{align} \label{eq:eq6}
\hat{W}_{j,\pm}^1 \ket{k}\ket{m} &= \left. \Big[ ( \sigma_x \otimes \ket{m}\bra{m} + \mathbb{I}\otimes \sum_{l \neq m} \ket{l}\bra{l} ) \right. \nonumber \\
& \left. S^1_{j,\pm} \left(\sigma_z \otimes \mathbb{I}^{\otimes n} \right) \right. \Big] \ket{k}\ket{m}.
\end{align} 

\noindent
{\it A Note on notation - } Here, uppercase letters are used to represent a particular qubit and lowercase letters refer to the order of the closed graph. It may also be observed from the Figs. \ref{fig:Scaling_even} and \ref{fig:Scaling_odd} that the $I^\text{th}$ qubit belongs to the closed graph of order $i = \frac{I}{2}$ if $I$ is even and $i = \frac{I-1}{2}$ if $I$ is odd.

In an abbreviated notation, the shift operator is written as,
\begin{equation}
\label{eq:eq7}
S^k_{j,\pm} = \sum_{l,m} \Big [ \ket{k}\bra{k} \otimes \ket{l_j\pm 1 \mod 4}\bra{l_j} + \ket{m\neq k}\bra{m}\otimes \mathbb{I}_p\Big ]. 
\end{equation}
\section{\label{sec:gates} Implementing Hadamard, Phase, and Controlled-NOT gates on n-qubit equivalent system} 
\noindent
\subsection*{{Hadamard Gate}} 

In this scheme, the Hadamard gate can be implemented on any qubit of $N$-qubit equivalent system  by redefining the Hadamard gates $\hat{H}_2$ and $\hat{H}_3$ in Ref.\,\cite{SPAC21}. Hadamard operation on the  $j^{th}$ level of the closed graph, when the coin state is $\ket{k}$  and position state is $\ket{m}$ as given by  Eq.\,\eqref{eq:eq4}, can be implemented on the quantum walk scheme by evolving the initial state by using Eq.\,\eqref{eq:eq8} when the Hadamard gate is applied on the $(2j)^\text{th}$-qubit and by using  Eq.\,\eqref{eq:eq9} when the Hadamard gate is applied on the $(2j+1)^\text{th}$-qubit.

\begin{widetext}
\begin{equation}
\label{eq:eq8}
	\hat{H}^k_{2,j} \ket{k}\ket{m} = \Big[ \hat{W}^{k \mod 2}_{j,-} \ket{0_j}\bra{0_j} + \hat{W}^{k \mod 2}_{j,+} \ket{1_j}\bra{1_j} + \hat{W}^{(k+1) \mod 2}_{j,-} \ket{3_j}\bra{3_j} + \hat{W}^{(k+1) \mod 2}_{j,+} \ket{2_j}\bra{2_j} \Big]  \left(\hat{H}_1 \otimes \mathbb{I}_2^{\otimes N}\right),
\end{equation}
\begin{equation}
\label{eq:eq9}
\hat{H}^k_{3,j}\ket{k}\ket{m} = \Big[ \hat{W}^{k \mod 2}_{j,+} \ket{0_j}\bra{0_j} + \hat{W}^{(k+1) \mod 2}_{j,-} \ket{1_j}\bra{1_j} + \hat{W}^{(k+1) \mod 2}_{j,+} \ket{3_j}\bra{3_j} + \hat{W}^{k \mod 2}_{j,-} \ket{2_j}\bra{2_j} \Big]   \left(\hat{H}_1 \otimes \mathbb{I}_2^{\otimes N}\right).
\end{equation}
\end{widetext}
Thus, the $\hat{H}$ corresponds to a position-dependent evolution operator in quantum walk scheme which applies to the appropriate vertices of the desired closed graph in the scaling diagram as shown in Figs. \ref{fig:Scaling_even} and \ref{fig:Scaling_odd}. Here the eigenstates of the 2-qubit equivalent $j^{th}$ closed system are $\ket{m_j}\, ,m=\{0,1,2,3\}$ .
The Hadamard on any qubit $Q > 1$ can be expressed on discrete-time quantum walk scheme in the form of evolution operator $\hat{H}^k_{i,j}$, where $i \in \{2,3\}$ and $j$ is the level of the closed graph such that  the relation between $j$ and $Q$ is  $j = \lfloor \frac{Q}{2} \rfloor$, i.e.,
\begin{equation}
\label{eq:eq10}
	\hat{H}_Q^k =
	\begin{cases}
		\hat{H}^k_{2,j} & \text{for even $Q$} \\
		\hat{H}^k_{3,j} & \text{for odd $Q$}.		
	\end{cases}
\end{equation}
A special case arises when the last qubit $Q=N$ is even and scaling is illustrated by Fig.\,\ref{fig:Scaling_even}. In this case, 
\begin{equation}
\label{eq:eq11}
	\hat{H}^k_Q = \hat{H}^k_{3,n}.
\end{equation}
In case $Q = 1$, the Hadamard gate can be reduced to a coin operation $\hat{H}_1 = \hat{C}\left(0,0,\frac{\pi}{4}\right)=\begin{bmatrix}1 & 1\\ 1 & -1\end{bmatrix}$ with an identity shift operator.

\noindent
\subsection*{ {Phase Gate} }
The Phase gate can be implemented on an $N$-qubit equivalent system in a manner similar to the Hadamard gate. Therefore, phase applied to the $Q^\text{th}$ qubit ($Q \in \{2,3,...N\}$) can be expressed in terms of the level $j$ of the closed graph as,
\begin{equation}
\label{eq:eq12}
	\hat{P}_Q =
	\begin{cases}
		\hat{P}_{2,j} & \text{for even $Q$} \\
		\hat{P}_{3,j} & \text{for odd $Q$} ,		
	\end{cases}
\end{equation}
where, $\hat{P}_{2,j}$ and $P_{3,j}$ are given as,
%
\begin{align} \label{eq:eq13}
P_{2,j} &= \mathbb{I}\otimes\left(\ket{0_j}\bra{0_j} + \ket{1_j}\bra{1_j}\right) + e^{i\phi}\mathbb{I}\otimes\left(\ket{3_j}\bra{3_j}+\ket{2_j}\bra{2_j}\right) \\
P_{3,j} &= \mathbb{I}\otimes\left(\ket{0_j}\bra{0_j} + \ket{2_j}\bra{2_j}\right) + e^{i\phi}\mathbb{I}\otimes\left(\ket{3_j}\bra{3_j}+\ket{1_j}\bra{1_j}\right)
\end{align}
%
For the special case when $Q=N$ is even, analogous to the Hadamard gate, phase gate can be given as,
\begin{equation}
\label{eq:eq14}
\hat{P}_N = \hat{P}_{3,n}.
\end{equation}
When $Q=1$, the phase operation on the first qubit is simply a coin operation, $C = \begin{bmatrix}1 & 0 \\ 0 & e^{i\phi}\end{bmatrix}$ with an identity operation on the position space.

\noindent
\subsection*{Controlled-NOT Gate }
Since, controlled-NOT gate (CNOT) is a two qubit gate (unlike Hadamard and phase gate), the gate implementation scheme changes form based on which two qubits are being addressed in the  $N$-qubit equivalent system. The different cases which will cover all the possibilities of controlled-NOT gate between control qubit $Q_c$ and target qubit $Q_t$ on $N$-qubit equivalent system are:
\subsubsection*{\bf Case 1: $Q_c=1$ or $Q_t=1$}
\noindent
{\it Case 1a: { $Q_c=1$, $Q_t$ is even, and $j=n$,} }
\begin{equation}
\label{eq:eq15}
	CNOT_{1,N} = \Big[ S^1_{j,+}\left(\ket{0_j}\bra{0_j}\right) + S^1_{j,-}\left(\ket{1_j}\bra{1_j}\right) \Big].
\end{equation}
{\it Case 1b: {$Q_t=1$, $Q_c$ is even, and  $i=n$,} }
\begin{equation}
\label{eq:eq16}
	CNOT_{N,1} = \Big[ \mathbb{I}\otimes \mathbb{I}\left(\ket{0_i}\bra{0_i}\right) + \sigma_{x}\otimes \mathbb{I}\left(\ket{1_i}\bra{1_i}\right) \Big].
\end{equation}
{\it Case 1c: $Q_c=1$, $Q_t$ is even, and is on $j^\text{th}$ level, with $j \neq n$,}
\begin{equation}
\label{eq:eq17}
\begin{split}
	CNOT_{1,Q_t} = \Big[ &S^1_{j,+}\left(\ket{1_j}\bra{1_j} + \ket{2_j}\bra{2_j}\right) \\ + &S^1_{j,-} \left(\ket{0_j}\bra{0_j} +\ket{3_j}\bra{3_j}\right) \Big].
\end{split}
\end{equation}
{\it Case 1d: $Q_c=1$, $Q_t$ is odd, and on the $j^\text{th}$-level for $j \neq n$,}
\begin{equation}
\label{eq:eq18}
\begin{split}
	CNOT_{1,Q_t} = \Big[ &S^1_{j,+}\left(\ket{0_j}\bra{0_j} + \ket{3_j}\bra{3_j}\right) \\ + &S^1_{j,-} \left(\ket{1_j}\bra{1_j} +\ket{2_j}\bra{2_j}\right) \Big].
\end{split}
\end{equation}
{\it Case 1e: $Q_t=1$, for even $Q_c$ such that $Q_c$ is on the $i^\text{th}$-level, and $i \neq n$,}
\begin{equation}
\label{eq:eq19}
\begin{split}
	CNOT_{Q_c,1} = \Big[ &\mathbb{I}\otimes \mathbb{I}\left(\ket{0_i}\bra{0_i} + \ket{1_i}\bra{1_i}\right)\\ + &\sigma_{x}\otimes \mathbb{I}\left(\ket{2_i}\bra{2_i} + \ket{3_i}\bra{3_i}\right) \Big].
\end{split}
\end{equation}
{\it Case 1f: $Q_t=1$, for odd $Q_c$ such that $Q_c$ is on the $i^\text{th}$-level and $i \neq n$,}
\begin{equation}
\label{eq:eq20}
\begin{split}
	CNOT_{Q_c,1} = \Big[ &\mathbb{I}\otimes \mathbb{I}\left(\ket{0_i}\bra{0_i} + \ket{2_i}\bra{2_i}\right)\\ + &\sigma_{x}\otimes \mathbb{I}\left(\ket{1_i}\bra{1_i} + \ket{3_i}\bra{3_i}\right) \Big].
\end{split}
\end{equation}

\subsubsection*{\bf Case 2: $Q_c$ and $Q_t$ are on the same level i.e., $i=j$. }
\noindent
{\it Case 2a: $Q_c$ is odd and $Q_t$ is even, }
\begin{equation}
\label{eq:eq21}
\begin{split}
CNOT_{Q_c,Q_t} = \Big[ &\mathbb{I}\otimes \mathbb{I}\left(\ket{0_j}\bra{0_j} + \ket{1_j}\bra{1_j}\right)\\ + &S^1_{j,+}S^0_{j,+}\left(\ket{2_j}\bra{2_j}\right) \\ + &S^1_{j,-}S^0_{j,-} \left(\ket{3_j}\bra{3_j}\right) \Big].
\end{split}
\end{equation}
{\it Case 2b: $Q_c$ is even and $Q_t$ is odd,}
\begin{equation}
\label{eq:eq22}
\begin{split}
CNOT_{Q_c,Q_t} = \Big[ &\mathbb{I}\otimes \mathbb{I}\left(\ket{0_j}\bra{0_j} + \ket{2_j}\bra{2_j}\right)\\ + &S^1_{j,+}S^0_{j,+}\left(\ket{1_j}\bra{1_j}\right) \\ + &S^1_{j,-}S^0_{j,-} \left(\ket{3_j}\bra{3_j}\right) \Big].
\end{split}
\end{equation}

\subsubsection*{\bf Case 3: $i \neq j$, where $Q_c$ and $Q_t$ are on $i^{th}$ and $j^{th}$ levels, respectively,  and $Q_t \neq N$ if $N$ is even}

\noindent
{\it Case 3a: $Q_c$ is odd and $Q_t$ is odd,}
\begin{equation}
\label{eq:eq23}
\small
\begin{split}
CNOT_{Q_c,Q_t} &= \Big[ \mathbb{I}\otimes \mathbb{I}\left(\ket{0_i}\bra{0_i} + \ket{2_i}\bra{2_i}\right)\\ + 
&S^1_{j,+}S^0_{j,+}\left(\ket{1_i}\bra{1_i} + \ket{3_i}\bra{3_i}\right)\left( \ket{0_j}\bra{0_j} + \ket{3_j}\bra{3_j} \right) \\ + 
&S^1_{j,-}S^0_{j,-} \left(\ket{1_i}\bra{1_i} + \ket{3_i}\bra{3_i}\right)\left( \ket{1_j}\bra{1_j} + \ket{2_j}\bra{2_j} \right) \Big].
\end{split}
\end{equation}
{\it Case 3b: $Q_c$ is odd and $Q_t$ is even,}
\begin{equation}
\label{eq:eq24}
\small
\begin{split}
CNOT_{Q_c, Q_t} &= \Big[ \mathbb{I}\otimes \mathbb{I}\left(\ket{0_i}\bra{0_i} + \ket{1_i}\bra{1_i}\right)\\ + 
&S^1_{j,+}S^0_{j,+}\left(\ket{2_i}\bra{2_i} + \ket{3_i}\bra{3_i}\right)\left( \ket{0_j}\bra{0_j} + \ket{3_j}\bra{3_j} \right) \\ + 
&S^1_{j,-}S^0_{j,-} \left(\ket{2_i}\bra{2_i} + \ket{3_i}\bra{3_i}\right)\left( \ket{1_j}\bra{1_j} + \ket{2_j}\bra{2_j} \right) \Big].
\end{split}
\end{equation}
{\it Case 3c: $Q_c$ and $Q_t$ are both even,}
\begin{equation}
\label{eq:eq25}
\small
\begin{split}
CNOT_{Q_c, Q_t} &= \Big[ \mathbb{I}\otimes \mathbb{I}\left(\ket{0_i}\bra{0_i} + \ket{1_i}\bra{1_i}\right)\\ + 
&S^1_{j,+}S^0_{j,+}\left(\ket{2_i}\bra{2_i} + \ket{3_i}\bra{3_i}\right)\left( \ket{1_j}\bra{1_j} + \ket{2_j}\bra{2_j} \right) \\ + 
&S^1_{j,-}S^0_{j,-} \left(\ket{2_i}\bra{2_i} + \ket{3_i}\bra{3_i}\right)\left( \ket{0_j}\bra{0_j} + \ket{3_j}\bra{3_j} \right) \Big].
\end{split}
\end{equation}
{\it Case 3d: $Q_c$ is even and $Q_t$ is odd,}\\
\begin{equation}
\small
\label{eq:eq26}
\begin{split}
CNOT_{Q_c, Q_t} &= \Big[ \mathbb{I}\otimes \mathbb{I}\left(\ket{0_i}\bra{0_i} + \ket{2_i}\bra{2_i}\right)\\ + 
&S^1_{j,+}S^0_{j,+}\left(\ket{1_i}\bra{1_i} + \ket{3_i}\bra{3_i}\right)\left( \ket{1_j}\bra{1_j} + \ket{2_j}\bra{2_j} \right) \\ + 
&S^1_{j,-}S^0_{j,-} \left(\ket{1_i}\bra{1_i} + \ket{3_i}\bra{3_i}\right)\left( \ket{0_j}\bra{0_j} + \ket{3_j}\bra{3_j} \right) \Big].
\end{split}
\normalsize
\end{equation}

\subsubsection*{\bf Case 4: $i \neq j$, where $Q_c$ and $Q_t$ are on $i^{th}$ and $j^{th}$level, respectively, {and $Q_t = N$, for even $N$}}
\noindent
{\it Case 4a: $Q_c$ is even,}
\begin{equation}
\label{eq:eq27}
\begin{split}
CNOT_{Q_c,Q_t} &= \Big[ \mathbb{I}\otimes \mathbb{I}\left(\ket{0_i}\bra{0_i} + \ket{1_i}\bra{1_i}\right)\\ + 
&S^1_{j,+}S^0_{j,+}\left(\ket{2_i}\bra{2_i} + \ket{3_i}\bra{3_i}\right)\left(\ket{0_j}\bra{0_j} \right) \\ + 
&S^1_{j,-}S^0_{j,-} \left(\ket{2_i}\bra{2_i} + \ket{3_i}\bra{3_i}\right)\left(\ket{1_j}\bra{1_j} \right) \Big].
\end{split}
\end{equation}
{\it Case 4b: $Q_c$ is odd,}
\begin{equation}
\label{eq:eq28}
\begin{split}
CNOT_{Q_c,Q_t} &= \Big[ \mathbb{I}\otimes \mathbb{I}\left(\ket{0_i}\bra{0_i} + \ket{2_i}\bra{2_i}\right)\\ + 
&S^1_{j,+}S^0_{j,+}\left(\ket{1_i}\bra{1_i} + \ket{3_i}\bra{3_i}\right)\left(\ket{0_j}\bra{0_j} \right) \\ + 
&S^1_{j,-}S^0_{j,-} \left(\ket{1_i}\bra{1_i} + \ket{3_i}\bra{3_i}\right)\left(\ket{1_j}\bra{1_j} \right) \Big].
\end{split}
\end{equation}
based on the two qubit on which CNOT gate is applied, different cases from above can be selected. Appendix \ref{apA} shows a different scheme of implementing the universal set of quantum gates on the same quantum walk scaling model. This shows that on this model of quantum walk, we can have different forms of the evolution operators to achieve desired operation based on the suitability of the available  quantum processors. This above scheme can be very easily implemented on photonic system with different sets of four-sited closed graph.  

\section{\label{sec:grover} Grover's Search algorithm on three qubit equivalent quantum walk scheme}
\begin{figure}
	\centering
	\includegraphics[width=0.9\linewidth]{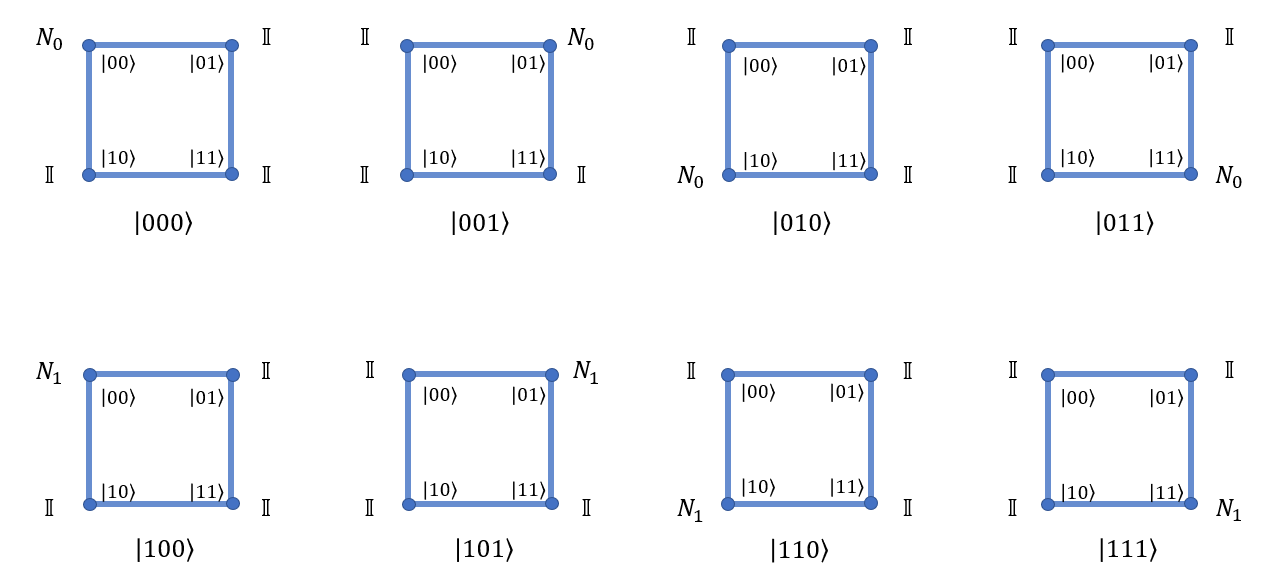}
	\caption{A schematic illustration of the oracle operation on the position state of the three-qubit equivalent quantum walk system using position dependent operators. The states below each square correspond to the target states of Grover's search. The definition of the various N operators have been defined in Eq.\,\eqref{eq:eq30}}
	\label{fig:GroverOracle}
\end{figure}

{ For searching a target state $\ket{x}$, Grover's search algorithm uses an oracle $\hat{\mathcal{O}}$ on state $\ket{\Psi} = \sum_{x} \psi_{x} \ket{x}$ of the form,
\begin{align}
\hat{\mathcal{O}}\ket{\Psi} \rightarrow \begin{cases}
-\ket{x} & \text{, when x is the target element} \\
\ket{x} & \text{, else} 
\end{cases}
\end{align}
 Grover's algorithm requires an oracle for the task of marking the targeted state by applying a negative sign to the desired search result state. 
 
The possible states of three-qubit system are $\ket{000}, \ket{001}, \ket{010}, \ket{011}, \ket{100}, \ket{101}, \ket{110}, \ket{111}$.} On three qubit equivalent quantum walk scheme, we need one real qubit on square lattice (closed graph of four sites). Oracle can be implemented by applying a position dependent evolution operator. The operator involves the coin operation, 
\begin{equation}
\label{eq:eq30}
\begin{split}
&\hat{N}_{1} =\sigma_{z} \otimes \mathbb{I} \\
&\hat{N}_{0} =\hat{C}(0,0,\pi) \otimes \mathbb{I}\\
&\hat{C}(\xi,\zeta,\theta)=\begin{bmatrix}
	e^{i\xi}\cos(\theta) & e^{i\zeta}\sin(\theta)\\
	e^{-i\zeta}\sin(\theta) & -e^{-i\xi}\cos(\theta)
	\end{bmatrix}.
\end{split}
\end{equation}
and the form of oracle on quantum walk scheme is shown in Fig.\,\ref{fig:GroverOracle}.

\begin{figure}
	\centering
	\includegraphics[width=0.6\linewidth]{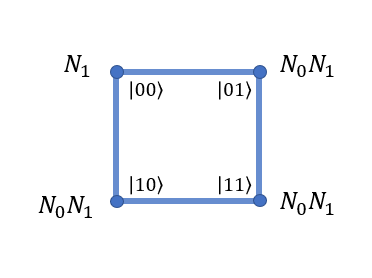}
	\caption{A schematic illustration of the iteration operation on the position basis of the three qubit system using position dependent operators. All the states except $\ket{000}$ will get a negative sign in this one step operation. The definition of the various N operators have been defined in Eq.\,\eqref{eq:eq30}}
	\label{fig:GroverIteration}
\end{figure}

Quantum walk scheme for three qubit Grover's search algorithm, when the coin and position state is initialized to $\ket{0}_c \otimes \ket{x=0}$, involves following steps,
\begin{enumerate}
	\item  A quantum walker starts with an equal superposition of all the states of the form $\ket{\psi_c} \otimes \ket{x}$ in both coin and position space. It can be achieved by applying operation $\hat{H}_2 \hat{H}_3$ on position state according to the quantum walk scheme as given in Ref.\,\cite{SPAC21} and then Hadamard operation on coin state.
	\item The oracle is applied on the walker according to the Fig.\,\ref{fig:GroverOracle} to search for the desired marked state.
	\item Hadamard operation is again applied on the coin state followed by the operation $\hat{H}_3 \hat{H}_2$ on position state according to quantum walk scheme. 
	\item The iteration method can be applied on the walker using position dependent $\hat{N}$ operators as defined in Eq.\,\eqref{eq:eq30} and illustrated in Fig.\,\ref{fig:GroverIteration} which will perform a conditional phase shift on every state except $\ket{000}$.
 \item Again apply Hadamard operation on the coin state followed by the operation $\hat{H}_3 \hat{H}_2$ on position state according to quantum walk scheme. 
	\item  Repeating steps 2 and 5 (also called the Grover iteration) for less or equal to $\lceil{\frac{\pi}{4}\sqrt{\frac{U}{V}}}\rceil$ times where, V = number of target entries in the search space and $U=2^N$. For $N=3$ and $V=1$, $\lceil{\frac{\pi}{4}\sqrt{\frac{U}{V}}}\rceil$ is $\leq 3 = 2$.
	\item Measurement in coin and position basis will give us our target state.
\end{enumerate}
Appendix \ref{apB} verifies the quantum walk based search algorithm by taking an example on search space of three qubit. 
\section{\label{sec:qft} Quantum Fourier Transformation on three-qubit equivalent quantum walk scheme}
The quantum Fourier transform is defined on orthonormal basis $\ket{0},\ket{1}...\ket{X-1}$ as a linear operator of the form,
 \begin{equation}
\label{eq:eq31}
    \ket{\alpha}=\frac{1}{\sqrt{X}}\sum_{l=0}^{X-1} e^{2 \pi i\alpha l/X}\ket{l}
\end{equation}
It can be transformed into a more easily implementable format as,
\begin{widetext}
\begin{align}
\ket{\alpha} &\xrightarrow{} \frac{1}{\sqrt{X}}\sum_{l=0}^{X-1} e^{2 \pi i\alpha l/X}\ket{l} \nonumber \\
&\xrightarrow{} \frac{1}{\sqrt{X}}(1\ket{0}+e^{2 \pi i0.\alpha_{N}}\ket{1})(1\ket{0}+e^{2 \pi i0.\alpha_{N-1} \alpha_{N}}\ket{1})... 
 (1\ket{0}+e^{2 \pi i0.\alpha_{1}...\alpha_{N-1} \alpha_{N}}\ket{1})\\
&[e^{2 \pi i \alpha_{1}....\alpha_{N-1}\alpha_{N}}=e^{2 \pi i0.\alpha_{N}}] \nonumber
\end{align}
\end{widetext}
where, $X=2^{N}$ and $N$ is the number of qubits in the system.
Quantum Fourier transformation on three-qubit quantum walk scheme requires a controlled-SWAP operation which, on quantum walk scheme can be obtained by applying the following operations,
\begin{figure}
    \centering
    \includegraphics[width=0.9\linewidth]{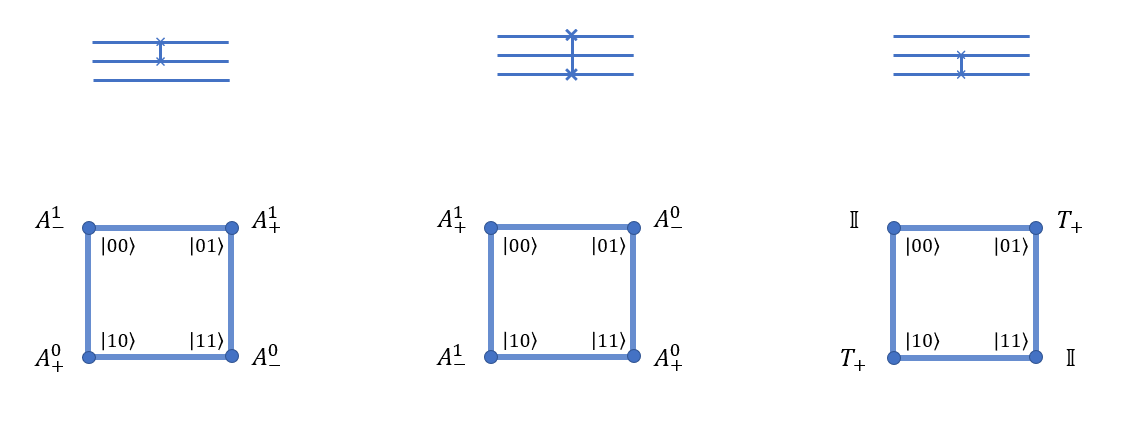}
    \caption{A schematic illustration of the controlled swap gate operation on the position basis of the three qubit equivalent quantum walk system using position dependent operators. The definition of the various A and T operators have been defined in Eq.\,\eqref{eq:eq32}}
    \label{fig:ControlledSwap}
\end{figure}
\begin{equation}
\label{eq:eq32}
    \begin{split}
        \hat{A}^{0}_{+}\ket{k,m} &=\hat{\sigma}_{x}^{m+1}\hat{S}_{1,+}^{0}\ket{k,m} \\
        \hat{A}^{1}_{+}\ket{k,m} &=\hat{\sigma}_{x}^{m+1}\hat{S}_{1,+}^{1}\ket{k,m}\\
        \hat{A}^{0}_{-}\ket{k,m} &=\hat{\sigma}_{x}^{m-1}\hat{S}_{1,-}^{0}\ket{k,m}\\
        \hat{A}^{1}_{-}\ket{k,m} &=\hat{\sigma}_{x}^{m-1}\hat{S}_{1,-}^{1}\ket{k,m}\\
        \hat{T}_{+}\ket{k,m}&=\hat{S}_{1,+}^1\hat{S}_{1,+}^1\hat{S}_{1,+}^0\hat{S}_{1,+}^0\ket{k,m};\\
    \end{split}
\end{equation}
where $\hat{S}^{k}_{1,\pm}$ are conditional shift operators in the position space of the walker and are given by  Eq.\,\eqref{eq:eq1} on the position state $\ket{m}$ conditioned on the state of coin $\ket{k}$ and $\hat{\sigma}^{m}_{x}$ is given by,
\begin{equation}
\label{eq:eq33}
        \hat{\sigma}^{m}_{x} =\hat{\sigma}_x \otimes (\ket{m}\bra{m})_p+\mathbb{I} \otimes \sum_{j\neq m}(\ket{j}\bra{j})
\end{equation}
Eqs.\,\eqref{eq:eq33} and \eqref{eq:eq32} and Fig.\,\ref{fig:ControlledSwap} outlines the operations which swaps two qubits.

\begin{figure}
    \centering
    \includegraphics[width=0.6\linewidth]{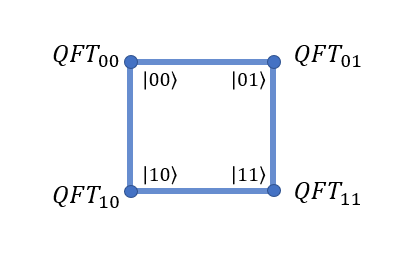}
    \caption{A schematic illustration of quantum Fourier transformation on three-qubit equivalent quantum walk scheme using position dependent operators.}
    \label{fig:FourierTransform}
\end{figure}

Thus, quantum Fourier transformation on quantum walk scheme can be given by the operation as shown in the Fig.\,\ref{fig:FourierTransform}, {after producing the initial state, where,
\begin{equation}
\label{eq:eq34}
\begin{split}
\mathcal{QFT}_{00} &= \hat{A}^{1}_{+} \hat{H}_3 \hat{H}_2 \hat{H}_1 \\
\mathcal{QFT}_{01} &= \hat{A}^{0}_{-} \hat{H}_3 \hat{H}_2 \hat{P}(\pi/4) \hat{H}_1  \\
\mathcal{QFT}_{11} &= \hat{A}^{0}_{+} \hat{H}_3 \hat{\Phi}(\pi/2) \hat{H}_2 \hat{P}(\pi/4) \hat{P}(\pi/2) \hat{H}_1 \\
\mathcal{QFT}_{10} &= \hat{A}^{1}_{-} \hat{H}_3 \hat{H}_2 \hat{\Phi}(\pi/2) \hat{H}_1 \\
\end{split}
\end{equation}
}
and operator $\hat{H}_2,\hat{H}_3$ and $\hat{P}(\phi), \hat{\Phi}(\phi) $ on the quantum walk scheme is given in the Ref.\,\cite{SPAC21}. $\hat{A}^{0}_{+,-}, \hat{A}^{1}_{+,-}$ are given by Eq.\,\eqref{eq:eq32} and $\hat{H}_1$ is Hadamard operation on coin operation. 

\section{\label{sec:qpe} Phase estimation algorithm on three qubit equivalent quantum walk scheme}
\begin{figure}
\centering
\includegraphics[width=0.95\linewidth]{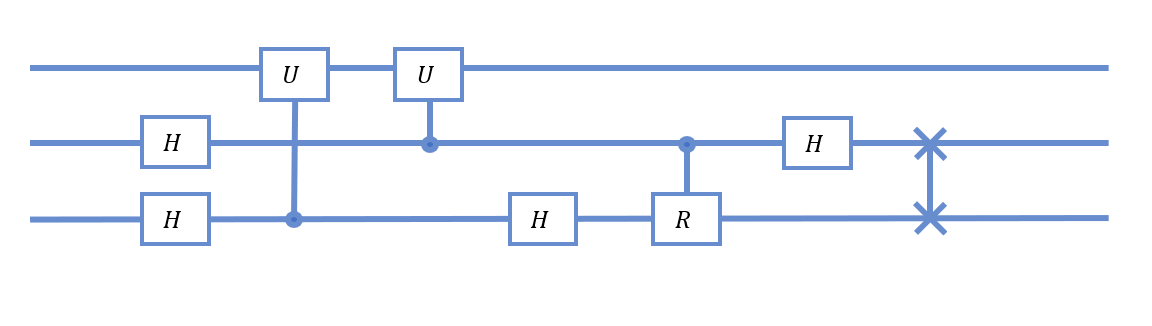}
\caption{Schematic of quantum circuit for phase estimation procedure on three qubit system. The state of the first qubit of the system is equivalent to the coin state and last two qubit shows the equivalence to the position states of the quantum walk scheme.}
\label{fig:PhaseEst}
\end{figure}

To estimate the phase $\varphi$ induced by an operator $\hat{U}$ on one of its eigenvectors $\ket{\psi}$ using single qubit on three-qubit equivalent quantum walk system, we consider the eigenvector $\ket{\psi}$ as the coin state and the position Hilbert space represents the state of the control qubits. The quantum circuit for phase estimation on three-qubit system is given in Fig.\,\ref{fig:PhaseEst}.

Algorithm for phase estimation on quantum walk scheme according to quantum circuit as given in Fig.\,\ref{fig:PhaseEst}, when coin and position state is initialised to state $\ket{0}_c \otimes \ket{x=0}$ is,
\begin{itemize}
\item[1.] Bringing the position states in equal superposition by implementing Hadamard operation $H_2$ and $H_3$ on second and third qubit, respectively. The state after this operation will have form,
\begin{equation}
\label{eq:eq35}
\begin{split}
\ket{\phi_1} &= \ket{0}_c \otimes \frac{\ket{00}+\ket{01}+\ket{10}+\ket{11}}{2} \\
&= \ket{0}_c \otimes \frac{\ket{x=0}+\ket{x=1}+\ket{x=3}+\ket{x=2}}{2}
\end{split}
\end{equation} 
\item[2.] Bringing the coin state to $\ket{\psi}_c$ using unitary operation $G$ such that $\ket{\psi}_c = G \ket{0}_c$. Here, $\ket{\psi}_c$ is an eigenvector of the unitary operator $U$ with eigenvalue $e^{2\pi i \varphi}$, where the value of $\varphi$ is unknown. 
The state after this operation will have form,
\begin{equation}
\label{eq:eq36}
\ket{\phi_2} = \ket{\psi}_c \otimes \frac{\ket{0}+\ket{1}+\ket{2}+\ket{3}}{2}
\end{equation} 
\item[3.] The effect of the controlled $\hat{U}$-operations can be thought of as different powers of $\hat{U}$ being operated on each of the position states as position-dependent coin operation given by,
\begin{equation}
\label{eq:eq37}
\begin{split}
    \hat{C}_U' &= \mathbb{I}_C \otimes \ket{0}\bra{0} + \hat{U} \otimes \ket{1}\bra{1}  \\
    &+ \hat{U}^2\otimes \ket{3}\bra{3} + U^3\otimes\ket{2}\bra{2}.
    \end{split}
\end{equation}
The form of the state after this operation is 
\begin{equation} 
\label{eq:eq38}
\begin{split}
    \ket{\phi_3} &= \hat{C}_U' \ket{\phi_2}  \\
    &= \frac{\ket{\psi}\ket{0}+\hat{U}\ket{\psi}\ket{1}+\hat{U}^2\ket{\psi}\ket{3}+\hat{U}^3\ket{\psi}\ket{2}}{2}  \\
    &= \ket{\psi} \otimes \frac{\ket{0}+e^{i\varphi}\ket{1}+e^{2i\varphi}\ket{3}+e^{3i\varphi}\ket{2}}{2}
    \end{split}
\end{equation}
\item[4.] Then applying inverse quantum Fourier transformation in the standard basis such that final state is,
\begin{equation}
\label{eq:eq39}
\begin{split}
\ket{\phi_{f}} &=  \mathcal{QFT}^{-1} \ket{\phi_3}  \\
&=\ket{\psi} \ket{\tilde{\varphi}}
\end{split}
\end{equation}
\end{itemize}
\noindent
The position dependent evolution operator for inverse Fourier transformation on state $\ket{\phi_3}$ in quantum walk scheme is given as,
\begin{equation}
\label{eq:eq40}
    \mathcal{QFT}^{-1} = (G \otimes \mathbb{I}) V_{2}^{x} V_{1}^{x} (G^{\dagger} \otimes \mathbb{I}), 
\end{equation}
where $G$ is the operator given in step-$2$ of the algorithm and the form of $V_{1}^{x}$ and $V_{2}^{x}$ position dependent operator is given as, 
\begin{equation}
\label{eq:eq41}
\begin{split}
V_{1}^{x=0} &= \hat{S}_{1}^{+}(\hat{H} \otimes \mathbb{I}) \\
V_{1}^{x=1} &= \hat{S}_{1}^{-}(\hat{H} \otimes \mathbb{I}) \\
V_{1}^{x=3} &= \hat{S}_{1}^{-}(\hat{H} \otimes \mathbb{I}) \\
V_{1}^{x=2} &= \hat{S}_{1}^{+}(\hat{H} \otimes \mathbb{I}) 
\end{split}
\end{equation}
and 
\begin{equation}
\label{eq:eq42}
\begin{split}
V_{2}^{x=0} &= \hat{S}_{1}^{-}(\hat{H} \otimes \mathbb{I})  \\
V_{2}^{x=1} &= \hat{S}_{1}^{+}(\hat{H} \otimes \mathbb{I}) (\hat{\Phi}_{-\frac{\pi}{2}}\hat{\sigma}_x\otimes\mathbb{I})  \\
V_{2}^{x=3} &= (\hat{\sigma}_x \otimes \mathbb{I}) \hat{S}_{0}^{+}(\hat{H} \otimes \mathbb{I})  \\
V_{2}^{x=2} &= (\hat{\sigma}_x \otimes \mathbb{I}) \hat{S}_{0}^{-}(\hat{H} \otimes \mathbb{I}) (\hat{\Phi}_{\frac{\pi}{2}}\hat{\sigma}_x\hat{\sigma}_z\otimes \mathbb{I}).
\end{split}
\end{equation}
Using this scheme on quantum walk, phase $\varphi$ induced by an operator $\hat{U}$ on one of its eigenvectors $\ket{\psi}_c$ can be estimated upto a certain accuracy. The accuracy in the estimation can be increased by using large position Hilbert space.

\section{\label{sec:complexity} Quantum Space and Time Complexity}

An analysis of complexity has its main concern regarding the inherent cost of solving a problem, where the cost is measured in terms of some well-defined resources. In this section, we shall be considering two ways of expressing complexity, namely {\it quantum space complexity}  and {\it quantum time complexity}. We define these terms as follows.

\begin{enumerate}
	\item {\bf Quantum space complexity} is defined as the number of real qubits required to implement the circuit.  This is analogous to the classical space complexity.
	
	\item {\bf Quantum time Complexity} is defined as the smallest number of time steps required to perform a computation on the circuit. In other words, it describes the least number of simultaneous elementary operations required to perform a single computation on the circuit. This is also in direct analogy to classical time complexity.

\end{enumerate}

In case of a standard circuit model, an elementary operation can be a single-qubit Hadamard gate, a single-qubit phase gate, or a two-qubit CNOT operation. Every other gate may be composed of these gates as they form a universal set \cite{AMC09}. 

In case of our model based on the quantum walk, an elementary operation is defined as a walk operation, i.e. a coin operation, followed by a shift operation. In case multiple quantum walk operations can be done with a common step, then the time complexity reduces. 

As an example, consider the sequence of steps $\hat{\Phi}({\frac{\pi}{2}})\hat{P}(\frac{\pi}{4})\hat{P}(\frac{\pi}{2})$, as used in the definition of $QFT_{11}$. With the way that $\Phi$ and $P$ gates are described in Ref. \cite{SPAC21}, both the gates can effectively be implemented by a coin operation, and can thus be combined into a single $P$ operation with a global phase. Thus, the time complexity of this 3 gate sequence is actually 1 time step.

Compared to the earlier universal quantum computation scheme with quantum walks\,\cite{LCE10}, our scheme defines computation purely as a sequence of walks that achieve the same effect as certain gates, instead of actually simulating gates from quantum walk steps, and then creating mirroring the circuit model. The existing models thus impose significant resource requirements to achieve the implementations
of algorithms, thereby becoming prohibitively resource-intensive.

We now detail an analysis of circuits for implementation of quantum algorithms considered in this paper, both in terms of the standard circuit model and our proposed quantum walk model of computation.

\subsection*{Grover's search} 

In this work, have considered Grover's search algorithm for 3 qubits, and have searched for the state $\ket{011}$ as an example.\\

\begin{figure}
	\centering
	\includegraphics[width=0.95\linewidth]{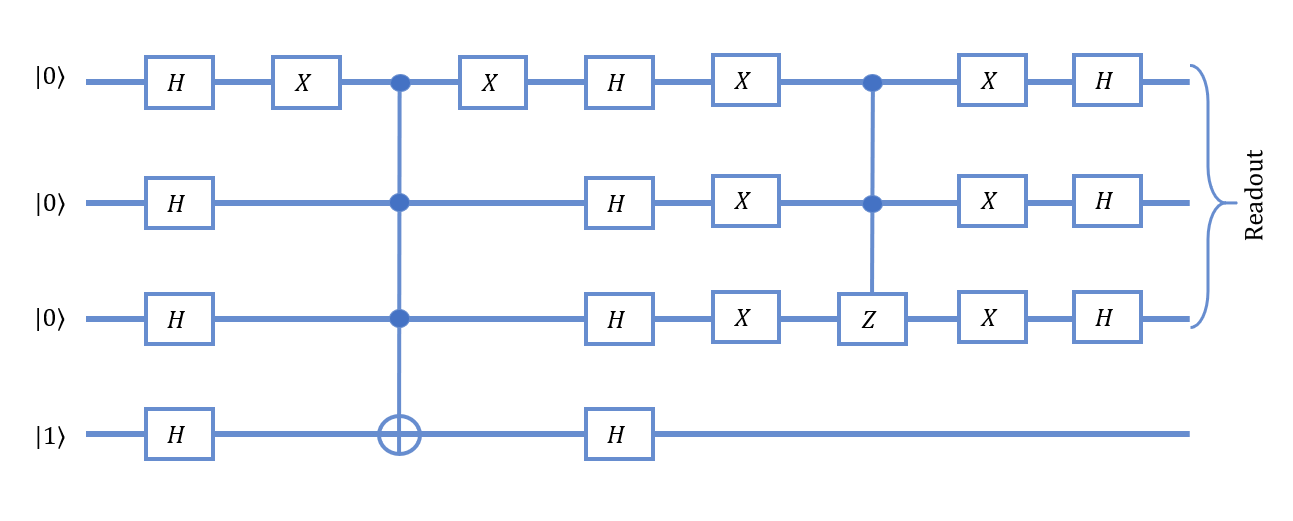}
	\caption{Schematic of quantum circuit for implementation of Grover's search algorithm on a three qubit system. The oracle is designed here to search for the state $|011\rangle$.}
	\label{fig:GroverCkt}
\end{figure}

\noindent
{\it Quantum space complexity}- The proposed quantum walk model of computation requires 3 qubits for implementation of the walk, however, only one qubit is a real (particle) qubit. The other two qubits are implemented with the position space. Thus, the quantum space complexity is 1. In case of the standard circuit model (Fig.\,\ref{fig:GroverCkt}), the implementation requires 3 qubits for the algorithm, and 1 ancilla qubit, thus making the total quantum space complexity to be 4.

\noindent
{\it Quantum time complexity}- In our quantum walk model, the generation of the initial superposition (done by the operator $H_2H_3$) takes 6 time steps. The oracle operation requires 1 time step, each ensuing Hadamard operation requires 3 time steps, and the final iteration operator needs another 2 time steps. Since 2 Grover iterations are required for a 3-qubit implementation, the total quantum time complexity becomes 39.

\begin{figure}
	\centering
	\includegraphics[width=0.95\linewidth]{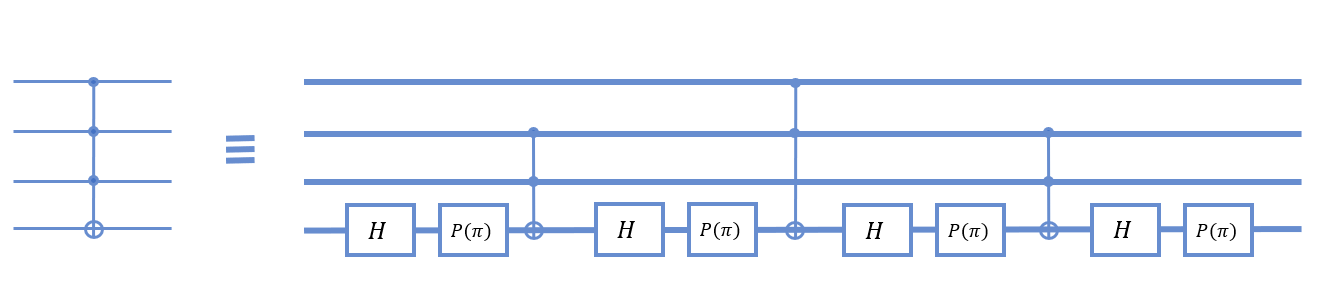}
	\caption{Schematic of quantum circuit for implementation of the $CCCNOT$ gate on a 4-qubit system. This gate has a quantum time complexity of 45.}
	\label{fig:CCCNOT}
\end{figure}

\begin{figure}
	\centering
	\includegraphics[width=0.95\linewidth]{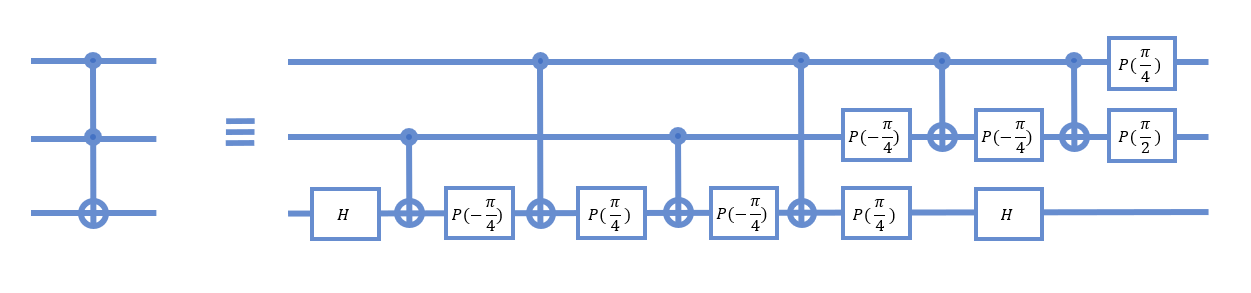}
	\caption{Schematic of quantum circuit for implementation of the Toffoli gate on a three qubit system. The quantum time complexity of this implementation is 13.}
	\label{fig:CCNOT}
\end{figure}

\begin{figure}
	\centering
	\includegraphics[width=0.95\linewidth]{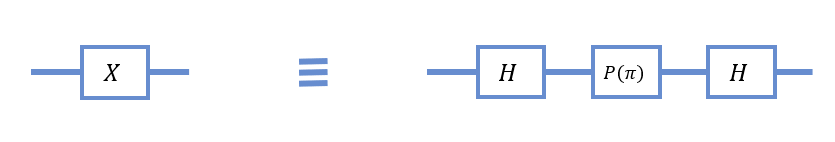}
	\caption{Schematic of quantum circuit for implementation of the Pauli X gate on a single qubit. This gate has a quantum time complexity of 3.}
	\label{fig:XGate}
\end{figure}

\begin{figure}
	\centering
	\includegraphics[width=0.95\linewidth]{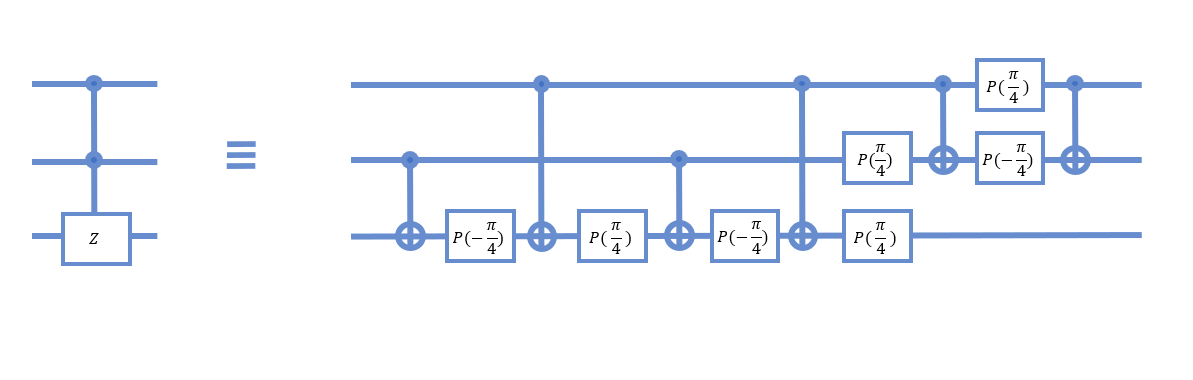}
	\caption{Schematic of quantum circuit for implementation of the $CCZ$ gate on a three qubit system. This implementation is similar to the 3-qubit Toffoli gate, except it has a few less operations. The quantum time complexity of this gate is thus 11.}
	\label{fig:CCZ}
\end{figure}

In the standard circuit, the superposition requires 4 parallel single-qubit gates on all 4 qubits and can be achieved in 1 time step. The various gates required to implement the algorithm on a 3-qubit system are the 4-qubit $CCCNOT$, which requires the Toffoli ($CCNOT$) gate implementation, the single qubit $X$ gate, and the $CCZ$ gate. The various gates and their quantum time complexities are shown in Figs. \ref{fig:CCCNOT}, \ref{fig:CCNOT}, \ref{fig:XGate}, and \ref{fig:CCZ}. Accounting for everything, the complete implementation has a quantum time complexity of 72.

\subsection*{Quantum Fourier Transform} 

We have considered the problem of computing the quantum Fourier transform for a 3-qubit system. \\

\noindent
{\it Quantum space complexity}- In our circuit, we require 1 real qubit to achieve a 3-qubit quantum Fourier transform. The standard circuit model requires 3 real qubits.\\

\noindent
{\it Quantum time complexity}- In our quantum walk-based model, the operations $A^{i}_{\pm}$ are essentially a single step of the walk, and can be implemented in one time step. The $A^{i}_{\pm}$ operation is then followed by the sequence $H_1H_3H_2$, which requires 7 time steps to implement.  The maximum time is required by $\mathcal{QFT}_{01}$ and $\mathcal{QFT}_{11}$ operators each of which require 9 time steps. This is due to the fact that the position-dependent Phase operations may be applied simultaneously, as they are all simply coin operations. Thus, the quantum-walk based model can implement this algorithm in 9 time steps.

\begin{figure}
	\centering
	\includegraphics[width=0.95\linewidth]{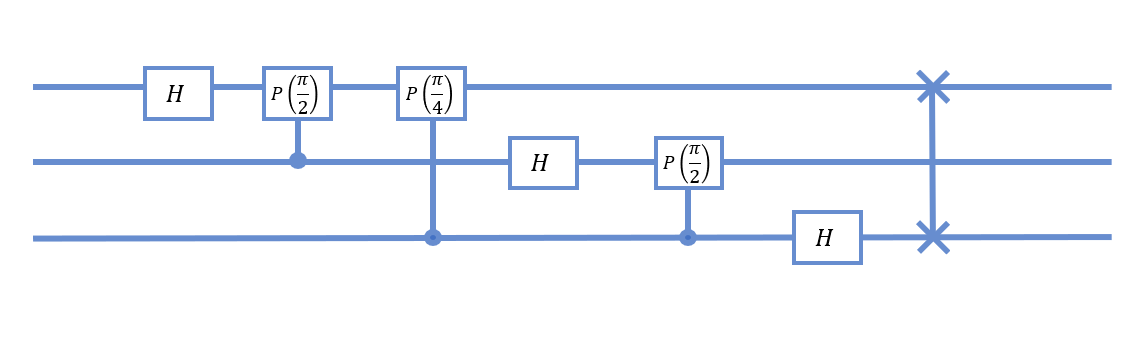}
	\caption{Schematic for the quantum circuit model implementation of the Quantum Fourier Transform on a three qubit system. The quantum time complexity of this implementation is 21.}
	\label{fig:QFTCkt}
\end{figure}

\begin{figure}
	\centering
	\includegraphics[width=0.95\linewidth]{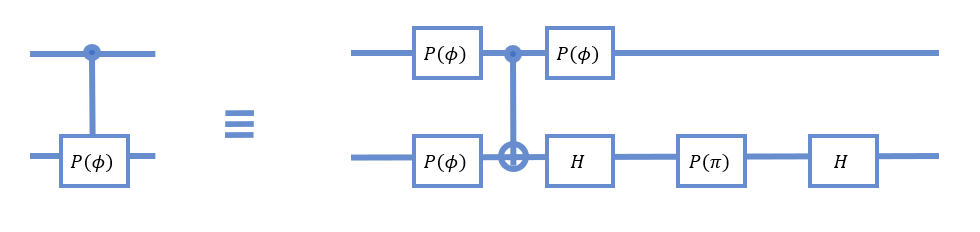}
	\caption{Schematic for the quantum circuit model implementation of the controlled Phase gate on two qubits. The quantum time complexity of this gate is 5.}
	\label{fig:CPhase}
\end{figure}

\begin{figure}
	\centering
	\includegraphics[width=0.95\linewidth]{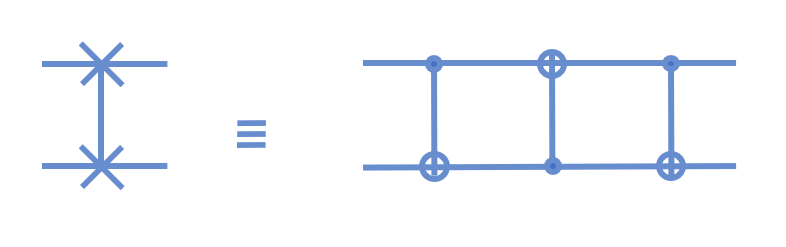}
	\caption{Schematic for the quantum circuit model implementation of the swap gate on two qubits. The quantum time complexity of this gate is 3.}
	\label{fig:2qubitSwap}
\end{figure}

In the standard circuit, the QFT is implemented as shown in the Fig. \ref{fig:QFTCkt}. The circuit begins with a Hadamard gate, followed by two controlled phase gates on the first qubit. The implementation of a controlled phase gate is shown in Fig. \ref{fig:CPhase}. From Fig. \ref{fig:CPhase}, it may be seen that a single controlled phase gate requires 5 time steps to implement. The final gate we require to implement is a two-qubit swap gate, which can be implemented efficiently as a series of 3 two-qubit CNOT gates, which requires 3 time steps to implement. The circuit is shown in Fig. \ref{fig:2qubitSwap} As a result, the standard circuit will require a total of 21 steps to implement.

\subsection*{Phase estimation algorithm}
We apply the phase estimation algorithm to an unknown unitary operation $U$. \\

\noindent
{\it Quantum space complexity}-In our circuit, we require only 1 real qubit in order to implement phase estimation. In a standard circuit, we need 3 real qubits to implement this algorithm.\\

\noindent
{\it Quantum time complexity}-In our circuit, the initial superposition required can be made in 6 time steps by the application of the operator $H_2H_3$. 1 time step is then required to implement the operator $G$, required to bring the coin into the correct state. It is sure that this will require only 1 time step as the coin qubit can be affected by a coin operator and an identity shift operator on the system. The controlled-$U$ operations are then realised as position-dependent operations, which require 3 time steps to implement (assuming $U$ will require 1 step to implement). The inverse Fourier transform on a 2-qubit system requires a worst case time of 7 steps. The complete quantum time complexity of this circuit thus becomes 17.

\begin{figure}
	\centering
	\includegraphics[width=\linewidth]{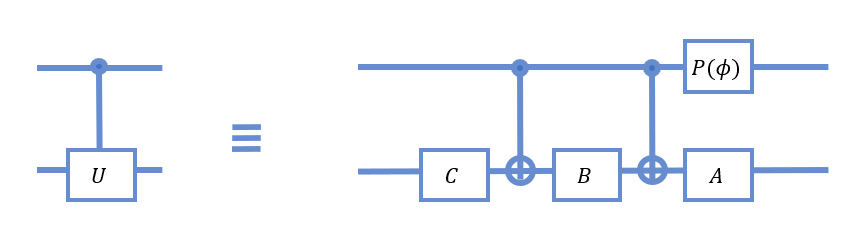}
	\caption{The circuit model of a controlled-$U$ gate, where $U$ is an unknown unitary, as given in ref.~\cite{NC10}. Here $P(\phi)$ represents the phase gate as described in \cite{SPAC21}, and $A,B,C,\phi$ satisfy $e^{i\phi}AXBXC = U$, and $ABC=\mathds{1}$. $X$ is the Pauli-$X$ operation.}
	\label{fig:CtrlU}
\end{figure}

In a standard circuit as shown in Fig. \ref{fig:PhaseEst}, the two initial Hadamard gates require one time step to implement, as they can be implemented in parallel. Going by the reduction for a controlled-$U$ gate, as shown in Fig.~\ref{fig:CtrlU}, the controlled-$U$ and controlled-$U^2$ gates would require 5 time steps each. The remaining circuit for an inverse QFT on two qubits requires 1 time step each for the Hadamard gates, 5 time steps for the controlled Phase, and 3 time steps for the swap gate. In total, the circuit requires 21 time steps to be implemented.

By this analysis, proposed quantum walk scheme uses a lesser number of real qubits to implement algorithmic operations than the circuit model. It also requires a lesser number of time steps than the circuit model in order to implement the algorithms shown here.

\section{\label{sec:error} Single-qubit error detection}

The proposed model of quantum computation also lends itself to an elementary representation of a quantum encoding. In this section, we present two examples of $[3,1]$ codes, and an example of a $[5,1]$ code. The $[3,1]$ code is able to detect either one of single-qubit errors, namely, bit-flip and phase-flip errors, and the $[5,1]$ code saturates the quantum Hamming bound, and is thus able to protect against arbitrary single-qubit errors.

\noindent
\subsection*{Bit-flip code}
\begin{figure}
	\includegraphics[width=\linewidth]{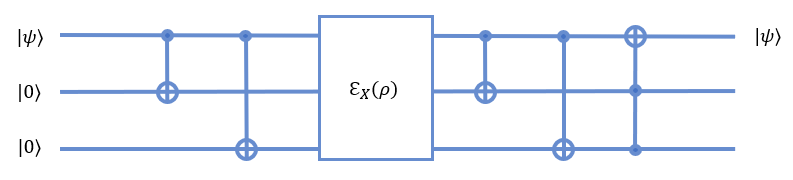}
	\caption{A circuit-model representation of the bit-flip code, implemented on a 3-qubit system. The figure is based on from the code as described in \cite{NC10}.}
	\label{fig:BitFlipCkt}
\end{figure}

The $[3,1]$ bit-flip encoding and decoding in the circuit model of computation is realised as shown in Fig~\ref{fig:BitFlipCkt}. The encoding uses 2 auxiliary qubits to generate error syndromes which can be corrected by the decoding circuit, which is shown post the introduction of error. The decoding of the syndrome and correction of error in a single qubit case requires the implementation of a Toffoli gate as shown. The Toffoli gate may be implemented with the gates belonging to the universal set as shown in Fig.~\ref{fig:CCNOT}.

The equivalent operation on a 3-qubit quantum walk system as detailed in \cite{SPAC21} may be performed by the operations $CNOT_{1,2}$ and $CNOT_{1,3}$ applied to the system. The final correction step is implemented with the Toffoli gate as demonstrated in Fig.~\ref{fig:BitFlipQW}.

\noindent
\subsection*{Phase-flip code}
\begin{figure}
	\includegraphics[width=\linewidth]{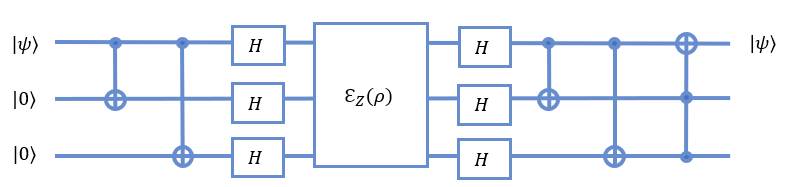}
	\caption{A circuit-model representation of the phase-flip code, implemented on a 3-qubit system. The figure is based on from the code as described in \cite{NC10}. This circuit is very similar to the bit-flip code, except that it requires an extra Hadamard operation on each qubit during both the detection and correction steps.}
	\label{fig:PhaseFlipCkt}
\end{figure}

The phase flip encoding is also a $[3,1]$ code, and is able to detect and correct single-qubit phase flip errors. The circuit representation for encoding and decoding in the phase flip code is shown in Fig.~\ref{fig:PhaseFlipCkt}. The circuit for the phase flip code is similar to that used for the bit flip encoding, except that it requires an extra Hadamard operation on each qubit after the bit-flip encoding. On a 3-qubit equivalent quantum walk system, this corresponds to applying the operations $H_1$, $H_2$, and $H_3$ on the system after applying the bit-flip encoding. 

\begin{minipage}[c]{\linewidth}
	\begin{tabular}{c}
		\includegraphics[width=0.9\textwidth]{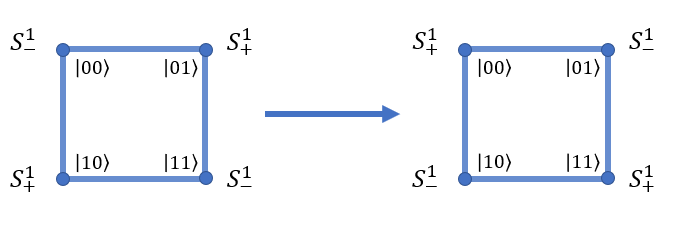} \\
		(a) \\
		\includegraphics[width=\textwidth]{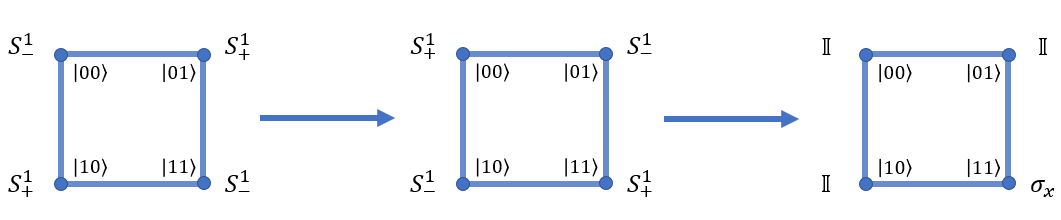} \\
		(b)
	\end{tabular}
	\captionof{figure}{A possible realization of the bit-flip encoding in the quantum walk paradigm. The figure (a) describes the steps in encoding, and (b) describes the decoding steps. The quantum time complexity of the complete encoding and decoding scheme is 5 (2 for encoding and 3 for decoding). In the circuit formalism, the quantum time complexity becomes 15 (1 for encoding, 14 for decoding). The reason for this disparity is that the quantum walk formalism allows for a simple realization of the Toffoli gate. \label{fig:BitFlipQW}}
\end{minipage}

\noindent
\subsection*{Error correcting code}
\begin{figure}
	\vspace*{5pt}
	\includegraphics[width=\linewidth]{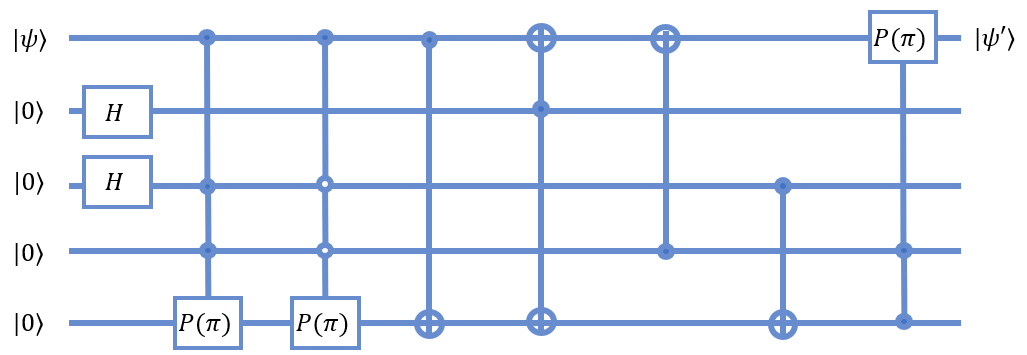}
	\caption{The quantum circuit for computing syndromes according to the $[5,1]$ code described in \cite{LMPZ96}. The circuit for recovery of the original qubit $\ket{\phi}$ is exactly the reverse of this circuit. The empty circle for the control qubit implies that the gate is activated if the qubit is in the $\ket{0}$ state. Gates are applied from the left column to the right column. Gates in a single column may be applied simultaneously.}
	\label{fig:PerfectQCodeCkt}
\end{figure}

Fig.~\ref{fig:PerfectQCodeCkt} shows a circuit model implementation of a $[5,1]$ encoding, a more elaborate description of which was given by  Laflamme \textit{et al.} \cite{LMPZ96}. The code enables error correction, and is able to correct against general single-qubit errors. This encoding may be implemented on a 2-level (5-qubit equivalent) graph in a quantum walk system, with one level consisting of a two-site closed graph and the second level being a four-site closed graph. This setup would require 2 real qubits to implement this code, however, in order to reduce the space complexity, it is possible to use a pair of 4-site closed graphs with a single particle executing a discrete time walk on them.

The quantum walk implementation would also require the implementation of the twin $CNOT$ gate, the controlled-controlled-Z ($CCZ$) gate, and the $CCCZ$ gate with two of the inputs inverted. The circuit model implementation of these gates is shown in Figs.~\ref{fig:CCZ} and \ref{fig:CCCZ}. The sequence of steps required to achieve the $CCZ$ gate on a quantum walk system is the same as illustrated in \cite{SPAC21}. The $CCCZ$ implementation will vary depending on the system topology chosen, however, it is illustrated here considering a 2-level implementation, where the both levels are 4-site graphs, traversed by a single particle. The forms of the operators required are illustrated in detail in Sec.~\ref{sec:gates}. The twin $CNOT$ gate may be designed in both the models in the same way, namely, by applying two $CNOT$ operations simultaneously.

\begin{widetext}
\begin{minipage}[c]{\linewidth}
	\includegraphics[width=0.145\linewidth]{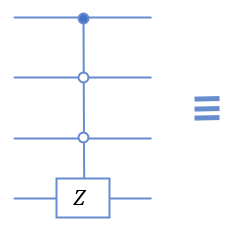} \hspace{6pt}
	\includegraphics[width=0.80\linewidth]{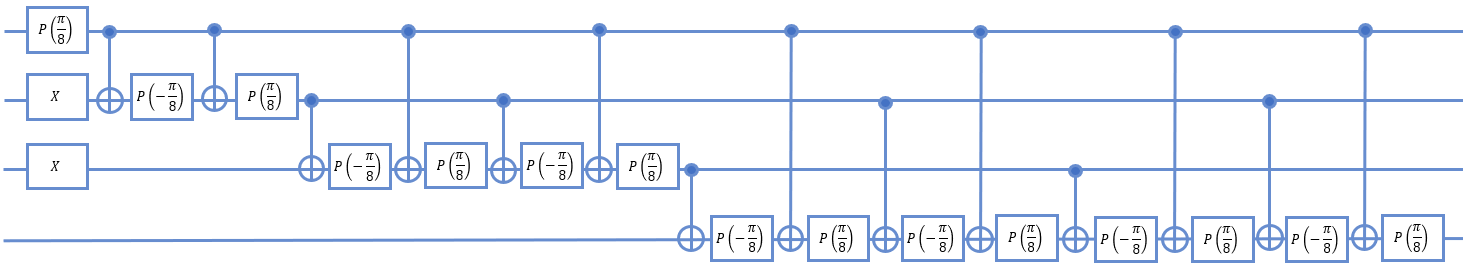}
	\captionof{figure}{A circuit-model realization of the modified $CCCZ$-gate as required for implementing the $[5,1]$ quantum error-correcting code. In order to realize the $CCCZ$-gate which activates as usual, i.e. when all inputs are $\ket{1}$, one may substitute the single qubit pauli $X$ rotations executed at the first time step with identity operations in this figure. The method to create this realization has been described in \cite{SS03}.}
	\label{fig:CCCZ}
\end{minipage}
\end{widetext}

\begin{figure}
	\includegraphics[width=0.5\linewidth]{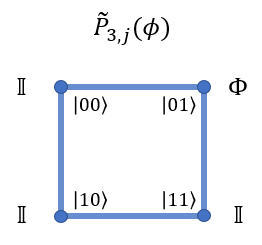}
	\caption{The modified form of the controlled-Phase operation in the quantum walk regime. This operator applies the phase when the control qubit is $\ket{0}$. The operation $\Phi$ adds a global phase, and is defined as described in~\cite{SPAC21}. }
		\label{fig:CPhaseQW}
\end{figure}

The $CCCZ$-gate requires a modified form of the controlled phase operation, which is given by the operator $\tilde{P}_{3,j}$, which will cause a conditional phase to be applied in case the control qubit is in the state $\ket{0}$. This is described in the Fig.~\ref{fig:CPhaseQW}. The form of the complete operation is given by Eq.~\ref{eq:eq44}.

\begin{widetext}
\begin{equation}
	\label{eq:eq44}
	CCCZ_{Q,\bar{b}, \bar{c},d} = \mathds{1}_Q \otimes \left( \ket{00}_{j=1} + \ket{10}_{j=1} \right) \otimes \tilde{P}_{3,j=2} + \mathds{1}_Q \otimes \left( \ket{01}_{j=1} + \ket{11}_{j=1} \right) \otimes \mathds{1}_{j=2},
\end{equation}
\end{widetext}

\noindent
where $j$ denotes the level at which the operation is applied, $Q$ is the first qubit in the system (assuming the qubit to be encoded is mapped to the real qubit), and the operation $\tilde{P}_{3,j=2}$ is as illustrated in Fig.~\ref{fig:CPhaseQW}.

\section{\label{sec:conc} Conclusion}

\noindent
In this paper, we have presented a more generalized form of the quantum computation using single particle quantum walk \cite{SPAC21}, and have shown the scaling of the scheme. Our proposed model can be scaled to system of a higher number of qubits by considering different sets of three-qubit equivalent closed graph as position space. To implement quantum universal gates on larger qubit equivalent system, the coin operation will control the evolution of the walker's position space by changing the probability amplitude of the targeted closed set. Using appropriate conditional position dependent evolution operators, the quantum walk based quantum computing scheme can be easily implemented.  We have also shown that on this scheme on an $N$-qubit system, universal gate implementation technique is not unique but can be changed according to the available resources. Since quantum walks on closed graph have been experimentally implemented on photonic system before \cite{XTA16,FYF18}, with the help of available photonic quantum processors, universal gates model based on single particle quantum walk can be implemented.

We have also presented the scheme for implementing quantum algorithms such as Grover's search, quantum Fourier transform  and quantum phase estimation on this scheme for three-qubit equivalent system. A comparison of circuit complexity and circuit depth shows that the proposed quantum walk scheme reduces the complexity when compared to circuit model in all of the cases. However, with a careful designing of position depended evolution operators one can engineer the implementation of various quantum computational tasks.

\begin{acknowledgments}
CMC would like to thank Department of Science and Technology,
Government of India for the Ramanujan Fellowship grant No.:SB/S2/RJN-192/2014.
We also acknowledge the support from Interdisciplinary Cyber Physical
Systems (ICPS) programme of the Department of Science and Technology,
India, Grant No.:DST/ICPS/QuST/Theme-1/2019/1. 
\end{acknowledgments}



\newpage

\appendix

\section{An alternative approach to scaling the DTQW to N qubits}
\label{apA}

The scheme presented for universal quantum computation using quantum walk for three qubit equivalent system as shown in \cite{SPAC21} can also be scaled to larger qubit system. It is done by using the position space of the preceding sets of the quantum walk system as the coin for the next set of quantum walk. Using the preceding set of walk as coin implies that the quantum walk is conditioned on the output of the preceding set of walk. Below we will show the scalability for four- and five-qubit systems and then extend it to $N$-qubit system. 

Form of shift operators which is used through out for scaling of the universal computation model as given in Ref.\,\cite{SPAC21} for input state $\ket{k,m,p}$ where, $\ket{k}$ is the coin state with two degree of freedom, $\ket{m}$ and $\ket{p}$ are the position states of two different cyclic quantum walk  with four state $span\{0,1,2,3\}$, respectively, is,
\begin{align}
\hat{W}_{\pm}^{0} &\equiv \big( \hat{\sigma_x} \otimes \ket{m}\bra{m} + 
\sum_{n \neq m} \mathds{1}_2 \otimes \ket{n}\bra{n} \big) \hat{S}_{2,\pm}^{k}\big(\hat{\sigma_{x}} \otimes \mathds{1} \big) , \nonumber \\
\hat{W}_{\pm}^{1} &\equiv \big( \hat{\sigma_x} \otimes \ket{m}\bra{m} + 
\sum_{n \neq m} \mathds{1}_2 \otimes \ket{n}\bra{n} \big) \hat{S}_{2,\pm}^{k}\big(\hat{\sigma_{z}} \otimes \mathds{1} \big)  ,
\end{align}
and
\begin{align}
\hat{V}_{\pm}^{0} &\equiv \big( \hat{\sigma_x}_2 \otimes \ket{p}\bra{p} + 
\sum_{q \neq p} \mathds{1}_4 \otimes \ket{q}\bra{q} \big) \hat{S}_{4,\pm}^{m}\big( \hat{\sigma_{x}}_{2} \otimes \mathds{1}  \big) \nonumber \\
&\equiv \hat{V}_{\pm}^{3} \nonumber \\
\hat{V}_{\pm}^{1} &\equiv \big( \hat{\sigma_x}_{2} \otimes \ket{p}\bra{p} + 
\sum_{q \neq p} \mathds{1}_4 \otimes \ket{q}\bra{q} \big) \hat{S}_{4,\pm}^{m}\big( \hat{\sigma_{z}}_{2} \otimes \mathds{1}  \big)  \nonumber \\
&\equiv \hat{V}_{\pm}^{2},
\end{align}
where, $\hat{\sigma_{x}}_2 =  \mathds{1}_2 \otimes \hat{\sigma_x}$ and on quantum walk system it is given as,
\begin{align} \label{sigma2}
\mathds{1}_2 \otimes \hat{\sigma_{x}}_2\ket{k,00} = \hat{S}_{2,+}^{0}\hat{S}_{2,+}^{1} \ket{k,0}\nonumber \\
\mathds{1}_2 \otimes \hat{\sigma_{x}}_2\ket{k,01} = \hat{S}_{2,-}^{0}\hat{S}_{2,-}^{1} \ket{k,1}\nonumber \\
\mathds{1}_2 \otimes \hat{\sigma_{x}}_2\ket{k,11} = \hat{S}_{2,+}^{0}\hat{S}_{2,+}^{1} \ket{k,2} \nonumber \\
\mathds{1}_2 \otimes \hat{\sigma_{x}}_2\ket{k,10} = \hat{S}_{2,-}^{0}\hat{S}_{2,-}^{1} \ket{k,3} .
\end{align}
If the coin state is 2-qubit equivalent then $\hat{\sigma_{x}}_2 =  \mathds{1}_2 \otimes \hat{\sigma_x}$ on quantum walk system it is given as,
\begin{align} \label{sigma4}
\mathds{1}_4 \otimes \hat{\sigma_{x}}_2\ket{m,00} = \hat{S}_{4,+}^{0}\hat{S}_{4,+}^{1} \hat{S}_{4,+}^{2}\hat{S}_{4,+}^{3} \ket{m,0}\nonumber \\
\mathds{1}_4 \otimes \hat{\sigma_{x}}_2\ket{m,01} = \hat{S}_{4,-}^{0}\hat{S}_{4,-}^{1} \hat{S}_{4,+}^{2}\hat{S}_{4,+}^{3} \ket{m,1}\nonumber \\
\mathds{1}_4 \otimes \hat{\sigma_{x}}_2\ket{m,11} = \hat{S}_{4,+}^{0}\hat{S}_{4,+}^{1} \hat{S}_{4,+}^{2}\hat{S}_{4,+}^{3} \ket{m,2} \nonumber \\
\mathds{1}_4 \otimes \hat{\sigma_{x}}_2\ket{m,10} = \hat{S}_{4,-}^{0}\hat{S}_{4,-}^{1} \hat{S}_{4,+}^{2}\hat{S}_{4,+}^{3} \ket{m,3} 
\end{align}
and $\hat{\sigma_{z}}_2 =  \mathds{1}_2 \otimes \hat{\sigma_z}$ and its action on a quantum walk system is defined by,
\begin{align}
\hat{\sigma_{z}}_2\ket{00} = \mathds{1}_2\ket{m=0}\nonumber \\
\hat{\sigma_{z}}_2\ket{01} = -\mathds{1}_2 \ket{m=1}\nonumber \\
\hat{\sigma_{z}}_2\ket{11} = -\mathds{1}_2 \ket{m=2} \nonumber \\
\hat{\sigma_{z}}_2\ket{10} = \mathds{1}_2 \ket{m= 3} 
\end{align}
The total number of states for this case will be equivalent to combined state of the Hilbert-space $\mathcal{H}_c \otimes \mathcal{H}_{p1} \otimes \mathcal{H}_{p2}$. 

\begin{widetext}
	
	\begin{align} \label{exampleHada}
	\Hat{H}_4 \ket{j}\otimes \ket{00} \otimes \ket{0} &\rightarrow (\mathds{1}_2 \otimes \hat{V}_{+}^{0}) (\mathbb{H}_3 \otimes \mathds{1}_2) (\ket{j,m=0,0}) \nonumber \\
	&= (\mathds{1}_2 \otimes \big( \hat{\sigma_x}_2 \otimes \ket{0}\bra{0} + 
	\mathds{1}_4 \otimes \ket{1}\bra{1} \big) \hat{S}_{4,+}^{0}\big( \hat{\sigma_{x}}_{2} \otimes \mathds{1}  \big)) (\hat{W}_{+}^{(j \mod 2)} \otimes \mathds{1}) (\hat{H}_1 \otimes \mathds{1} \otimes \mathds{1}) (\ket{j,m=0,0}) \nonumber \\
	&= (\mathds{1}_2 \otimes \big( \hat{\sigma_x}_2 \otimes \ket{0}\bra{0} + 
	\mathds{1}_4 \otimes \ket{1}\bra{1} \big) \hat{S}_{4,+}^{0}\big( \hat{\sigma_{x}}_{2} \otimes \mathds{1}  \big)) \frac{1}{\sqrt{2}}(\ket{j,m=0,0} + \ket{j,m=1,0}) \nonumber \\
	&= (\mathds{1}_2 \otimes \big( \hat{\sigma_x}_2 \otimes \ket{0}\bra{0} + 
	\mathds{1}_4 \otimes \ket{1}\bra{1} \big) \hat{S}_{4,+}^{0})\frac{1}{\sqrt{2}}(\ket{j,m=1,0} + \ket{j,m=0,0}) \nonumber \\
	&= (\mathds{1}_2 \otimes \big( \hat{\sigma_x}_2 \otimes \ket{0}\bra{0} + 
	\mathds{1}_4 \otimes \ket{1}\bra{1} \big))\frac{1}{\sqrt{2}}(\ket{j,m=1,0} + \ket{j,m=0,1}) \nonumber \\
	&= \frac{1}{\sqrt{2}}(\ket{j,m=0,0} + \ket{j,m=0,1}) \nonumber \\
	&= \ket{j}\otimes \ket{00} \otimes \frac{\Big(\ket{0} +  \ket{1}\Big)}{\sqrt{2}}
	\end{align}
	
\end{widetext}

\subsection{\label{apA1} Implementing Hadamard, phase, and CNOT-gate on four qubit equivalent system} 
\noindent
{\it Hadamard operation:} To map the Hadamard operation on fourth qubit of the four qubit system, one can use a three-qubit equivalent quantum walk system and a set of position space with two states such that the operation is defined on $\mathcal{H}_c \otimes \mathcal{H}_{4} \otimes \mathcal{H}_{2}$ combined Hilbert space. The three-qubit equivalent quantum walk system defined on $\mathcal{H}_c \otimes \mathcal{H}_{4}$ will act as a coin for the next set of position space with two states $span\{\ket{0},\ket{1}\}$ defined on $\mathcal{H}_{2}$ . One can realize the Hadamard operation on the fourth qubit of the four-qubit system by using a combination of coin and shift operators as,
\begin{align} \label{H_4}
\hat{H}_4\ket{j,m,0} &\rightarrow  (\mathds{1}_2 \otimes \hat{V}_{+}^{(m \mod 4)}) (\hat{H}_3 \otimes \mathds{1}_2) \ket{j,m,0}, \nonumber \\
\hat{H}_4\ket{j,m,1} &\rightarrow (\mathds{1}_2 \otimes \hat{V}_{-}^{(m+1 \mod 4)}) (\hat{H}_3 \otimes \mathds{1}_2) \ket{j,m,1}
\end{align}
%
where, $\hat{H}_3$ is given by the quantum walk scheme presented in the Ref.\,\cite{SPAC21}. $\ket{j}$ is the basis state of coin Hilbert space such that $\ket{j} = \{\ket{0}, \ket{1}\}$, $\ket{l}$ represent two qubit equivalent cyclic position Hilbert state $\mathcal{H}_4$ given by, $\ket{m} = \{\ket{x=0}, \ket{x=1}, \ket{x=2}, \ket{x=3}\} \equiv \{\ket{00}, \ket{01}, \ket{11}, \ket{10}\}$. $\hat{H}_4 \equiv \mathds{1} \otimes \mathds{1} \otimes \mathds{1} \otimes  \hat{H}$ and an illustration of this scaling when $\ket{m=0}$ is given in Eq.\,\eqref{exampleHada}.
\noindent
{\it Phase operation:} To map the phase gate operation on fourth qubit of the four qubit system, one can again use a three-qubit equivalent quantum walk system as coin for the position space with two states $span\{\ket{0}, \ket{1}\}$. The walk is again defined on the Hilbert space $\mathcal{H}_c \otimes \mathcal{H}_{4} \otimes \mathcal{H}_{2}$. The operations that will evolve the initial state of the quantum walk into the state with phase on the fourth qubit is,
\begin{align}
\hat{P}_4\ket{j,m,0} &\rightarrow \mathds{1}_2 \otimes \mathds{1}_4 \otimes \mathds{1}_2 \ket{j,m,0} \nonumber \\
\hat{P}_4 \ket{j,m,1} &\rightarrow \hat{\Phi} \otimes \mathds{1}_4 \otimes \mathds{1}_2 \ket{j,m,1}
\end{align} 

where, $\hat{\Phi} = e^{i\phi}\mathds{1}_2$ and $\hat{P}_4 \equiv \mathds{1} \otimes \mathds{1} \otimes \mathds{1} \otimes  \hat{P}$.
\noindent
{\it Controlled-NOT operation:} To map the CNOT gate operation when the fourth qubit is the target and other qubits are control of the four qubit system, we will again need a three-qubit equivalent quantum walk system as coin for the next set of position space $span\{\ket{0}, \ket{1}\}$. The quantum walk scheme when fourth qubit is the target for various control qubit $CNOT_{i4}$, where $i$ is the control qubit, is given by,
\begin{align}
\hat{CNOT}_{14} \ket{j,m,0} &\rightarrow (\hat{S}_{2,-}^{0} \otimes \mathds{1}_2) (\mathds{1}_2 \otimes \hat{S}_{4,+}^{m}) (\hat{S}_{2,+}^{0} \otimes \mathds{1}_2)  \nonumber \\
\hat{CNOT}_{14} \ket{j,m,1} &\rightarrow (\hat{S}_{2,-}^{0} \otimes \mathds{1}_2) (\mathds{1}_2 \otimes\hat{S}_{4,-}^{m})(\hat{S}_{2,+}^{0} \otimes \mathds{1}_2);
\end{align}
\begin{align}
\hat{CNOT}_{24} \ket{j,m,0} &\rightarrow (\mathds{1}_2 \otimes \hat{S}_{4,+}^3) (\mathds{1}_2 \otimes \hat{S}_{4,+}^2)  \nonumber \\
\hat{CNOT}_{24} \ket{j,m,1} &\rightarrow (\mathds{1}_2 \otimes \hat{S}_{4,-}^3) (\mathds{1}_2 \otimes \hat{S}_{4,-}^2) 
\end{align}
and
\begin{align}
\hat{CNOT}_{34} \ket{j,m,0} &\rightarrow (\mathds{1}_2 \otimes \hat{S}_{4,+}^1) (\mathds{1}_2 \otimes \hat{S}_{4,+}^2)   \nonumber \\
\hat{CNOT}_{34} \ket{j,m,1} &\rightarrow (\mathds{1}_2 \otimes \hat{S}_{4,-}^1) (\mathds{1}_2 \otimes \hat{S}_{4,-}^2) 
\end{align}
The quantum walk scheme when fourth qubit is the control for various target qubit $CNOT_{4i}$, where $i$ is the target qubit and is given by,
\begin{align}
\hat{CNOT}_{41} \ket{j,m,0} &\rightarrow \mathds{1} \nonumber \\
\hat{CNOT}_{41} \ket{j,m,1} &\rightarrow (\hat{\sigma}_x \otimes \mathds{1}_4 \otimes \mathds{1}_2);
\end{align}
\begin{align}
\hat{CNOT}_{42} \ket{j,m,0} &\rightarrow \mathds{1}  \nonumber \\
\hat{CNOT}_{42} \ket{j,m,1} &\rightarrow (\mathds{1}_2 \otimes \hat{\sigma}_{x1}) 
\end{align}
where, $\hat{\sigma_{x}}_1 =  \hat{\sigma_x} \otimes \mathds{1}_2$ and on quantum walk system it is given as,
\begin{align} \label{sigma1}
\mathds{1}_2 \otimes \hat{\sigma_{x}}_1\ket{k,00} = \hat{S}_{2,-}^{0}\hat{S}_{2,-}^{1} \ket{k,0}\nonumber \\
\mathds{1}_2 \otimes \hat{\sigma_{x}}_1\ket{k,01} = \hat{S}_{2,+}^{0}\hat{S}_{2,+}^{1} \ket{k,1}\nonumber \\
\mathds{1}_2 \otimes \hat{\sigma_{x}}_1\ket{k,11} = \hat{S}_{2,-}^{0}\hat{S}_{2,-}^{1} \ket{k,2} \nonumber \\
\mathds{1}_2 \otimes \hat{\sigma_{x}}_1\ket{k,10} = \hat{S}_{2,+}^{0}\hat{S}_{2,+}^{1} \ket{k,3} 
\end{align} 
and
\begin{align}
\hat{CNOT}_{43} \ket{j,m,0} &\rightarrow \mathds{1}   \nonumber \\
\hat{CNOT}_{43} \ket{j,m,1} &\rightarrow (\mathds{1}_2 \otimes \hat{\sigma}_{x2})
\end{align}
where, $\hat{\sigma_{x}}_2 =  \mathds{1}_2 \otimes \hat{\sigma_x}$ on quantum walk system is given by Eq.\,\eqref{sigma2}.

\subsection{\label{apA2} Implementing Hadamard, phase, and Controlled-NOT operation on five qubit equivalent system} 
\noindent
{\it Hadamard Operation:} The Hadamard operation on fourth qubit of the five qubit system is defined on two cyclic quantum walk system with four position states. The combined Hilbert space is $\mathcal{H}_c \otimes \mathcal{H}_{4} \otimes \mathcal{H}_{4}$ . Position state of the previous quantum walk system will act as a coin for the position space of the next quantum walk system. One can realize the Hadamard operation on the fourth qubit of the five-qubit system by using a combination of coin and shift operators as,
\begin{align} \label{H_54}
\hat{H}_4\ket{j,m,00} &\rightarrow  (\mathds{1}_2 \otimes \hat{V}_{-}^{(m \mod 4)}) (\hat{H}_3 \otimes \mathds{1}_4) \ket{j,m,p=0}, \nonumber \\
\hat{H}_4\ket{j,m,01} &\rightarrow (\mathds{1}_2 \otimes \hat{V}_{+}^{(m \mod 4)}) (\hat{H}_3 \otimes \mathds{1}_4) \ket{j,m,p=1}, \nonumber \\
\hat{H}_4\ket{j,m,11} &\rightarrow  (\mathds{1}_2 \otimes \hat{V}_{-}^{(m+1 \mod 4)}) (\hat{H}_3 \otimes \mathds{1}_4) \ket{j,m,p=2}, \nonumber \\
\hat{H}_4\ket{j,m,10} &\rightarrow (\mathds{1}_2 \otimes \hat{V}_{+}^{(m+1 \mod 4)}) (\hat{H}_3 \otimes \mathds{1}_4) \ket{j,m,p=3}
\end{align}
and
\begin{align} \label{H_55}
\hat{H}_5\ket{j,m,00} &\rightarrow  (\mathds{1}_2 \otimes \hat{V}_{+}^{(m \mod 4)}) (\hat{H}_3 \otimes \mathds{1}_4) \ket{j,m,p=0}, \nonumber \\
\hat{H}_5\ket{j,m,01} &\rightarrow (\mathds{1}_2 \otimes \hat{V}_{-}^{(m+1 \mod 4)}) (\hat{H}_3 \otimes \mathds{1}_4) \ket{j,m,p=1}, \nonumber \\
\hat{H}_5\ket{j,m,11} &\rightarrow  (\mathds{1}_2 \otimes \hat{V}_{+}^{(m+1 \mod 4)}) (\hat{H}_3 \otimes \mathds{1}_4) \ket{j,m,p=2}, \nonumber \\
\hat{H}_5\ket{j,m,10} &\rightarrow (\mathds{1}_2 \otimes \hat{V}_{-}^{(m \mod 4)}) (\hat{H}_3 \otimes \mathds{1}_4) \ket{j,m,p=3}.
\end{align}
Here $\hat{H}_3$ is again given in the Ref.\,\cite{SPAC21}, $\hat{H}_4 \equiv \mathds{1} \otimes \mathds{1} \otimes \mathds{1} \otimes  \hat{H} \otimes \mathds{1}$ and $\hat{H}_5 \equiv \mathds{1} \otimes \mathds{1} \otimes \mathds{1} \otimes \mathds{1} \otimes \hat{H}$ .
\noindent
{\it Phase operation:} Similarly, the quantum walk scheme for the phase operation on fourth qubit of the five qubit system is also defined on the Hilbert space $\mathcal{H}_c \otimes \mathcal{H}_{4} \otimes \mathcal{H}_{4}$. The operations that will evolve the initial state of the quantum walk into the state with phase on the fourth qubit is,
\begin{align}
\hat{P}_4\ket{j,m,00} &\rightarrow \mathds{1} \ket{j,m,p=0}, \nonumber \\
\hat{P}_4\ket{j,m,01} &\rightarrow \mathds{1} \ket{j,m,p=1}, \nonumber \\
\hat{P}_4\ket{j,m,11} &\rightarrow (\hat{\Phi} \otimes \mathds{1}) \ket{j,m,p=2}, \nonumber \\
\hat{P}_4\ket{j,m,10} &\rightarrow (\hat{\Phi} \otimes \mathds{1}) \ket{j,m,p=3}.
\end{align} 
and phase operation on the fifth qubit is,
\begin{align}
\hat{P}_5\ket{j,m,00} &\rightarrow \mathds{1} \ket{j,m,p=0}, \nonumber \\
\hat{P}_5\ket{j,m,01} &\rightarrow (\hat{\Phi} \otimes \mathds{1}) \ket{j,m,p=1}, \nonumber \\
\hat{P}_5\ket{j,m,11} &\rightarrow (\hat{\Phi} \otimes \mathds{1}) \ket{j,m,p=2}, \nonumber \\
\hat{P}_5\ket{j,m,10} &\rightarrow \mathds{1}  \ket{j,m,p=3}.
\end{align} 
where, $\hat{\Phi} = e^{i\phi}\mathds{1}_2$, $\hat{P}_4 \equiv \mathds{1} \otimes \mathds{1} \otimes \mathds{1} \otimes  \hat{P} \otimes \mathds{1}$ and $\hat{P}_5 \equiv \mathds{1} \otimes \mathds{1} \otimes \mathds{1} \otimes \mathds{1} \otimes  \hat{P}$.

{\it Controlled-NOT operation:} To implement the CNOT operation when the fourth and fifth qubits are the target of the five qubit system, we will need a cyclic quantum walk with four position basis state as a coin for the position space of another quantum walk on the first system with four position basis states. The quantum walk scheme when fourth qubit is the target for various control qubit $CNOT_{i4}$, where $i$ is the control qubits, is given by,
\begin{align}
\hat{CNOT}_{14} \ket{j,m,00} &\rightarrow (\hat{S}_{2,-}^{0} \otimes \mathds{1}_2) (\mathds{1}_2 \otimes \hat{S}_{4,-}^{m}) (\hat{S}_{2,+}^{0} \otimes \mathds{1}_2)  \nonumber \\
\hat{CNOT}_{14} \ket{j,m,01} &\rightarrow (\hat{S}_{2,-}^{0} \otimes \mathds{1}_2) (\mathds{1}_2 \otimes\hat{S}_{4,+}^{m})(\hat{S}_{2,+}^{0} \otimes \mathds{1}_2) \nonumber \\
\hat{CNOT}_{14} \ket{j,m,11} &\rightarrow (\hat{S}_{2,-}^{0} \otimes \mathds{1}_2) (\mathds{1}_2 \otimes \hat{S}_{4,-}^{m}) (\hat{S}_{2,+}^{0} \otimes \mathds{1}_2)  \nonumber \\
\hat{CNOT}_{14} \ket{j,m,10} &\rightarrow (\hat{S}_{2,-}^{0} \otimes \mathds{1}_2) (\mathds{1}_2 \otimes\hat{S}_{4,+}^{m})(\hat{S}_{2,+}^{0} \otimes \mathds{1}_2) ;
\end{align}
\begin{align}
\hat{CNOT}_{24} \ket{j,m,00} &\rightarrow (\mathds{1}_2 \otimes \hat{S}_{4,-}^3) (\mathds{1}_2 \otimes \hat{S}_{4,-}^2)  \nonumber \\
\hat{CNOT}_{24} \ket{j,m,01} &\rightarrow (\mathds{1}_2 \otimes \hat{S}_{4,+}^3) (\mathds{1}_2 \otimes \hat{S}_{4,+}^2) \nonumber \\
\hat{CNOT}_{24} \ket{j,m,11} &\rightarrow (\mathds{1}_2 \otimes \hat{S}_{4,-}^3) (\mathds{1}_2 \otimes \hat{S}_{4,-}^2)  \nonumber \\
\hat{CNOT}_{24} \ket{j,m,10} &\rightarrow (\mathds{1}_2 \otimes \hat{S}_{4,+}^3) (\mathds{1}_2 \otimes \hat{S}_{4,+}^2);
\end{align}
\begin{align}
\hat{CNOT}_{34} \ket{j,m,00} &\rightarrow (\mathds{1}_2 \otimes \hat{S}_{4,-}^1) (\mathds{1}_2 \otimes \hat{S}_{4,-}^2)   \nonumber \\
\hat{CNOT}_{34} \ket{j,m,01} &\rightarrow (\mathds{1}_2 \otimes \hat{S}_{4,+}^1) (\mathds{1}_2 \otimes \hat{S}_{4,+}^2) \nonumber \\
\hat{CNOT}_{34} \ket{j,m,11} &\rightarrow (\mathds{1}_2 \otimes \hat{S}_{4,-}^1) (\mathds{1}_2 \otimes \hat{S}_{4,-}^2)   \nonumber \\
\hat{CNOT}_{34} \ket{j,m,10} &\rightarrow (\mathds{1}_2 \otimes \hat{S}_{4,+}^1) (\mathds{1}_2 \otimes \hat{S}_{4,+}^2)
\end{align}
and
\begin{align}
\hat{CNOT}_{54} \ket{j,m,00} &\rightarrow \mathds{1}   \nonumber \\
\hat{CNOT}_{54} \ket{j,m,01} &\rightarrow (\mathds{1}_2 \otimes \hat{S}_{4,+}^m) \nonumber \\
\hat{CNOT}_{54} \ket{j,m,11} &\rightarrow (\mathds{1}_2 \otimes \hat{S}_{4,-}^m)    \nonumber \\
\hat{CNOT}_{54} \ket{j,m,10} &\rightarrow \mathds{1}
\end{align}
Similarly, by having another combinations of the shift operators $\hat{S}_{2,\pm}$ and $\hat{S}_{4,\pm}$  Eq.\,\eqref{eq:eq1} and coin operators Eq.\,\eqref{eq:eq2}, one can easily implement the CNOT operation on fifth qubit as both target $CNOT_{i5}$ or control $CNOT_{5i}$ here, $i-$ is the control or target qubit, respectively. 

Universal computation on the $(n-1)^\text{th}$ and $n^\text{th}$ qubit of the n-qubit system using quantum walk scheme when $n$ is odd number, will require $(n-1)/2$ sets of quantum walk with four position basis states. The walk is defined on combined Hilbert space $\mathcal{H}_c \otimes \mathcal{H}_4 \otimes ... \otimes \mathcal{H}_4$ as shown in Fig.\,\ref{fig:Scaling_odd}. Similarly, if $n$ is even in n-qubit system, it will require $(n/2)-1$ sets of quantum walk with four position basis states and one set of quantum walk with two position basis states. Here the walk will be defined on Hilbert space $\mathcal{H}_c \otimes \mathcal{H}_4 \otimes ... \otimes \mathcal{H}_2$ as shown in Fig.\,\ref{fig:Scaling_even}. $\mathcal{H}_c$ is the coin Hilbert space with two internal states $\{\ket{0},\ket{1}\}$ which acts as the coin for the position Hilbert space $\mathcal{H}_4$ of the first set of the cyclic quantum walk with four computational basis states $\{\ket{0}, \ket{1}, \ket{2}, \ket{3}\}$ equivalent to $\{\ket{00}, \ket{01}, \ket{11}, \ket{10}\}$, respectively. The position space of the first set of cyclic quantum-walk will act as coin for the next set of quantum walk with four position basis states and so on. This scheme can be scaled to $n$- qubit system by using the position space of the previous set of quantum-walk as coin for the position space of next set of the quantum walk.

\begin{widetext}
	
	\begin{table} 
		\centering
		\caption{Hadamard operation $\hat{H}$ on even $(n-1)^\text{th}$  and odd $n^\text{th}$ qubit when the processor has $n$ number of qubits. $\ket{m}$ is the position basis state of the previous set of cyclic quantum walk. $\ket{(n-1)} \otimes \ket{n}$ is $\{\ket{00}, \ket{01}, \ket{11}, \ket{10}\}$ which is also equivalent to the four computational position basis state of the cyclic quantum walk.}
		\begin{tabular}{ |c|c|c|c| } 
			\hline
			& $\hat{H}_{(n-1)}$ & $\hat{H}_{n}$ \\
			\hline
			$\ket{j,...,m,00}$ & $(\mathds{1} \otimes \hat{V}_{-}^{(m \mod 4)}) (\hat{H}_{n-2} \otimes \mathds{1}_4) \ket{j,...,m,p=0} $& $(\mathds{1} \otimes \hat{V}_{+}^{(m \mod 4)}) (\hat{H}_{n-2} \otimes \mathds{1}_4)\ket{j,...,m,p=0}$ \\ 
			\hline
			$\ket{j,...,m,01}$ & $(\mathds{1} \otimes \hat{V}_{+}^{(m \mod 4)}) (\hat{H}_{n-2} \otimes \mathds{1}_4)\ket{j,...,m,p=1}$ & $(\mathds{1} \otimes \hat{V}_{-}^{(m+1 \mod 4)}) (\hat{H}_{n-2} \otimes \mathds{1}_4)\ket{j,...,m,p=1}$ \\
			\hline 
			$\ket{j,...,m,11}$ & $(\mathds{1} \otimes \hat{V}_{-}^{(m+1 \mod 4)}) (\hat{H}_{n-2} \otimes \mathds{1}_4)\ket{j,...,m,p=2} $ & $(\mathds{1} \otimes \hat{V}_{+}^{(m+1 \mod 4)}) (\hat{H}_{n-2} \otimes \mathds{1}_4) \ket{j,...,m,p=2}$ \\
			\hline
			$\ket{j,...,m,10}$ & $(\mathds{1} \otimes \hat{V}_{+}^{(m+1 \mod 4)}) (\hat{H}_{n-2} \otimes \mathds{1}_4)\ket{j,...,m,p=3}$ & $(\mathds{1} \otimes \hat{V}_{-}^{(m \mod 4)}) (\hat{H}_{n-2} \otimes \mathds{1}_4)\ket{j,...,m,p=3}$ \\
			\hline
		\end{tabular}
		\label{tab:tableHada1}
	\end{table}
	
	\begin{table} 
		\centering
		\caption{Phase operation $\hat{P}$ on even $(n-1)^\text{th}$  and odd $n^\text{th}$ qubit when the processor has $n$ number of qubits. $\ket{m}$ is the position basis states of the previous set of cyclic quantum walk. $\ket{(n-1)} \otimes \ket{n}$ is $\{\ket{00}, \ket{01}, \ket{11}, \ket{10}\}$ which is also equivalent to the four computational position basis states of the cyclic quantum walk.}
		\begin{tabular}{ |c|c|c|c| } 
			\hline
			& $\hat{P}_{(n-1)}$ & $\hat{P}_{n}$ \\
			\hline
			$\ket{j,...,m,00}$ & $ \mathds{1} \ket{j,...,m,p=0} $& $ \mathds{1} \ket{j,...,m,p=0}$ \\ 
			\hline
			$\ket{j,...,m,01}$ & $\mathds{1} \ket{j,...,m,p=1}$ & $(\hat{\Phi} \otimes \mathds{1}) \ket{j,...,m,p=1}$ \\
			\hline 
			$\ket{j,...,m,11}$ & $(\hat{\Phi} \otimes \mathds{1}) \ket{j,...,m,p=2} $ & $(\hat{\Phi} \otimes \mathds{1}) \ket{j,...,m,p=2}$ \\
			\hline
			$\ket{j,...,m,10}$ & $(\hat{\Phi} \otimes \mathds{1}) \ket{j,...,m,p=3}$ & $\mathds{1} \ket{j,...,m,p=3}$ \\
			\hline
		\end{tabular}
		\label{tab:tablePhase1}
	\end{table}
	
\end{widetext}
\par
\noindent
{\it Hadamard operation:} Generalised scheme of quantum walk computation to implement Hadamard operation on $(n-1)^\text{th}$ and $n^\text{th}$ qubit of the n-qubit system when $n$ is odd is given in the table\,\ref{tab:tableHada1}. An illustration of the scaling of the quantum walk scheme when the number of qubit in the system is odd is given in Fig.\,\ref{fig:Scaling_odd}. Quantum walk scheme illustration is shown in Fig.\,\ref{fig:Scaling_even} when the number of qubit in the system is even and to implement Hadamard operation on the last qubit $\ket{l}$ when the number of qubits in the system is even is given by,
\begin{equation}
\label{eq:eqA1}
\begin{split}
\hat{H}_l\ket{j,...,m,0} &= (\mathds{1} \otimes \hat{V}_{+}^{(m \mod 4)}) (\hat{H}_{p-1} \otimes \mathds{1}_2) \ket{j,...,m,0}   \\
\hat{H}_l\ket{j,...,m,1} &= (\mathds{1} \otimes \hat{V}_{-}^{(m+1 \mod 4)}) (\hat{H}_{p-1} \otimes \mathds{1}_2) \ket{j,...,m,1}
\end{split}
\end{equation}
\noindent
{\it Phase operation:} Phase operation on $(n-1)^\text{th}$ and $n^\text{th}$ qubit of the n-qubit system when $n$ is odd in number is given in the table\,\ref{tab:tablePhase1} on quantum walk scheme. Similar to Hadamard operation, phase operation on the last qubit $\ket{l}$ when the number of qubit in the system is even is given by,
\begin{align}
\hat{P}_l\ket{j,...,m,0} &= \mathds{1} \ket{j,...,m,0}\ket{j,...,m,0}  \nonumber \\
\hat{P}_l\ket{j,...,m,1} &= (\hat{\Phi} \otimes \mathds{1}) \ket{j,...,m,1}
\end{align}
An illustration of the scaling of the quantum walk scheme when the number of the qubit in the system is odd and even is shown in Figs.\,\ref{fig:Scaling_odd} and \ref{fig:Scaling_even}, respectively.

\noindent
{\it Controlled-NOT operation:} CNOT operation can be implemented between any two qubits using quantum walk scheme with the help of the coin and shift operators given in the Eqs.\,\eqref{eq:eq1} and \eqref{eq:eq2} along with identity operation in a similar way as CNOT operation has been shown for four and five qubit system. The quantum walk scheme will need same setup as shown in Figs.\,\ref{fig:Scaling_even} and \ref{fig:Scaling_odd} for a system with even and odd number of qubits.

\section{An illustration of 3-qubit Grover's Search Algorithm with a DTQW} \label{apB}

{\it An example of the quantum walk based search algorithm,} on search space of three qubit system with the state $\ket{011}$ marked as our target state is presented below.
\begin{enumerate}
	\item We start with a state $\ket{\psi}=\frac{1}{\sqrt{2}}(\ket{0}+\ket{1})_c\frac{1}{\sqrt{4}}(\ket{00}+\ket{01}+\ket{10}+\ket{11})_p$ 
	\begin{align}
	\ket{\psi}&=\sqrt{\frac{1}{8}}(\ket{0}_c\ket{00}_p+\ket{0}_c\ket{01}_p+\ket{0}_c\ket{10}_p +\ket{0}_c\ket{11}_p \nonumber \\
	&+\ket{1}_c\ket{00}_p+\ket{1}_c\ket{01}_p+\ket{1}_c\ket{10}_p +\ket{1}_c\ket{11}_p)\\
	&=\left(\frac{\cos{\frac{\theta}{2}}}{\sqrt{7}}(\ket{0}_c\ket{00}_p+\ket{0}_c\ket{01}_p +\ket{0}_c\ket{10}_p +\ket{1}_c\ket{00}_p \right. \nonumber \\
	& \left. +\ket{1}_c\ket{01}_p+\ket{1}_c\ket{10}_p+\ket{1}_c\ket{11}_p)  + \sin{\frac{\theta}{2}}(\ket{0}_c\ket{11}_p)\right)
	\label{eqn:Pn31}
	\end{align}
	where $\cos{\frac{\theta}{2}}=\sqrt{\frac{7}{8}}$ and $\sin{\frac{\theta}{2}}=\sqrt{\frac{1}{8}}$.
	
	\item Now we apply the oracle on this $\ket{\psi}$. The oracle for target state $\ket{011}$ is represented by the following operation
	\begin{equation}
	\hat{O}=\mathds{1}(\ket{00}\bra{00}+\ket{10}\bra{10}+\ket{01}\bra{01})_p + N_0 (\ket{11}\bra{11})_p
	\label{eqn:Pn32}
	\end{equation}
	where the definition of N operators is given in Eq.\,\ref{eq:eq30}.
	The above operation applies identity operator on $\ket{00},\ket{10},\ket{01}$ positions states and hence the probabilities in these position states are left untouched. Lets see the operation on $\ket{11}$ position state :
	
	Term $\ket{0}_c\ket{11}_p$:
	\begin{align*}
	\hat{O}\ket{0}_c\ket{11}_p&=N_0\ket{0}_c\ket{11}_p\\
	&=\hat{C}(0,0,\pi) \otimes \mathds{1}\ket{0}_c\ket{11}_p\\
	&=\begin{vmatrix}
	-1 & 0\\
	0 & 1
	\end{vmatrix}\begin{vmatrix}
	1\\
	0
	\end{vmatrix} \otimes \mathds{1} \ket{11}_p\\
	&=-\ket{0}_c\ket{11}_p\\
	\end{align*}

	Term $\ket{1}_c\ket{11}_p$:
	\begin{align*}
	\hat{O}\ket{1}_c\ket{11}_p&=N_0\ket{1}_c\ket{11}_p\\
	&=\hat{C}(0,0,\pi) \otimes \mathds{1}\ket{1}_c\ket{11}_p\\
	&=\begin{vmatrix}
	-1 & 0\\
	0 & 1
	\end{vmatrix}\begin{vmatrix}
	0\\
	1
	\end{vmatrix} \otimes \mathds{1} \ket{11}_p\\
	&=\ket{1}_c\ket{11}_p\\
	\end{align*}
	
	So the final state after oracle operation is
	\begin{align}
	\ket{\psi}'&=\left(\frac{\cos{\frac{\theta}{2}}}{\sqrt{7}}(\ket{0}_c\ket{00}_p+\ket{0}_c\ket{01}_p +\ket{0}_c\ket{10}_p+\ket{1}_c\ket{00}_p \right. \nonumber \\
	& \left. +\ket{1}_c\ket{01}_p+\ket{1}_c\ket{10}_p+\ket{1}_c\ket{11}_p) - \sin{\frac{\theta}{2}}(\ket{0}_c\ket{11}_p)\right).
	\label{eqn:Pn33}
	\end{align}
	
	\item Similar to step 2 above, the operation in Fig.\,\ref{fig:GroverIteration} gives all states except $\ket{0}_c\ket{00}_p$ a negative sign and this along with the Hadamard operation gives the following state :
	\begin{align}
	\ket{\psi}''&=\left(\frac{\cos{\frac{3\theta}{2}}}{\sqrt{7}}(\ket{0}_c\ket{00}_p+\ket{0}_c\ket{01}_p+\ket{0}_c\ket{10}_p +\ket{1}_c\ket{00}_p \right. \nonumber \\
	& \left. +\ket{1}_c\ket{01}_p+\ket{1}_c\ket{10}_p+\ket{1}_c\ket{11}_p) - \sin{\frac{3\theta}{2}}(\ket{0}_c\ket{11}_p)\right).
	\label{eqn:Pn34}
	\end{align}
	
	\item We need to perform the Grover iteration(step 2 and 3) two ($CI(\frac{\arccos{\sqrt{\frac{1}{8}}}}{2\arccos{\sqrt{\frac{7}{8}}}})$) times and the last state would be
	\begin{align}
	\ket{\psi}'''&=\left(\frac{\cos{\frac{5\theta}{2}}}{\sqrt{7}}(\ket{0}_c\ket{00}_p+\ket{0}_c\ket{01}_p+\ket{0}_c\ket{10}_p+\ket{1}_c\ket{00}_p \right. \nonumber \\
	& \left.+\ket{1}_c\ket{01}_p+\ket{1}_c\ket{10}_p+\ket{1}_c\ket{11}_p)- \sin{\frac{5\theta}{2}}(\ket{0}_c\ket{11}_p)\right).
	\label{eqn:Pn35}
	\end{align}
	So our target state's probability is $|\sin^2{\frac{5\theta}{2}}|^2=0.945$ given $\sin{\frac{\theta}{2}}=\frac{1}{\sqrt{8}}$. Hence, after the above operations, if we do a measurement in coin and position basis we will get the target state with a very high probability. One major advantage is that, this scheme does not require ancilla qubits and both the oracle and iteration operations are just position dependent coin operations.
\end{enumerate}

\end{document}